\documentstyle[prd,aps,preprint,epsfig]{revtex}
\firstfigfalse

\begin{document}

\draft

\preprint{\rightline{ANL-HEP-PR-02-005}}

\newcommand{\Dirac}{\rlap{\hspace{-.5mm} \slash} D}
\newcommand{\sumint}{\rlap{\hspace{-.5mm} $\sum$} \int}
\title{ 
 The Phase Diagram of Four Flavor SU(2) Lattice Gauge Theory at Nonzero
Chemical Potential and Temperature} 
\vskip -1 truecm

\author{John~B.~Kogut and Dominique Toublan }
\address {Physics Department, University of Illinois at Urbana-Champaign,
Urbana, IL 61801-3080}

\author{D.~K.~Sinclair.}
\address{HEP Division, Argonne National Laboratory, 9700
        South Cass Avenue, Argonne, IL 60439, USA}

\date{\today}
\maketitle
\vskip -1 truecm

\begin{abstract}
$SU(2)$ lattice gauge theory with four flavors of quarks is simulated at nonzero
chemical potential $\mu$ and temperature $T$ and the results are
compared to the predictions of Effective Lagrangians. Simulations
on $16^4$ lattices indicate that at zero $T$ the theory experiences a second
order phase transition to a diquark condensate state. Several methods of analysis, including
equation of state fits suggested by Chiral Perturbation Theory, suggest that
mean-field scaling describes this critical point.
Nonzero $T$ and $\mu$ are studied on $12^3 \times 6$ lattices. For
low $T$, increasing $\mu$ takes the system through a line of second order
phase transitions to a diquark condensed phase. Increasing $T$ at high $\mu$,
the system passes through a line of first order transitions from the diquark
phase to the quark-gluon plasma phase. Metastability is found in the vicinity
of the first order line. There is a tricritical point along this line of
transitions whose position is consistent with theoretical predictions.
\end{abstract}


\newpage

\section{Introduction}

Recently there has been a resurgence of interest in QCD at nonzero chemical
potential for quark number. Arguments based on instantons \cite{Shuryak} and
phenomenological gap equations \cite{Wilczek} support the old idea that
diquark condensation and a color superconductivity phase transition
\cite{Love,Barrois} will occur at a critical chemical potential somewhat
greater than that at which nuclear matter forms. Nuclear matter is
expected to be the ground state of QCD above a chemical potential
$\mu_N$, which is slightly 
less than one-third of the proton's mass. These arguments neglect the forces of
confinement and may only be reliable at asymptotically large chemical
potentials. Since diquarks carry color in real QCD, a quantitative
understanding of confinement and screening is needed to estimate the energies
of the states which control the phases of the system. Unfortunately, brute
force lattice simulation methods, which are so useful at $T\neq0$ and $\mu=0$, 
do not yet provide a reliable simulation algorithm for these environments in
which the fermion determinant becomes complex.

Given these problems, theorists have turned to simpler models which can be
analyzed and simulated by current methods. One of the more interesting is the
color $SU(2)$ version of QCD which addresses some of the issues of interest
\cite{Shuryak}, \cite{Wilczek}, \cite{Hands}. In this model diquarks do not
carry color, so their condensation does not break color symmetry dynamically.
Instead, the diquark condensed phase resembles a superfluid rather than a
superconductor. The critical chemical potential is also expected to be
one-half the mass of the lightest meson, the pion, because quarks and
anti-quarks reside in equivalent representations of the $SU(2)$ color group.
Hence hadron flavor multiplets include both mesons and diquarks.
Chiral Lagrangians can be used to study the diquark condensation transition in
this model because the critical chemical potential vanishes in the chiral
limit, and the model has a Goldstone realization of the spontaneously broken
quark-number symmetry \cite{Toublan,SUNY,STV1,SSS}. The problem has
also been 
studied within a Random Matrix Model at non-zero $\mu$ and $T$
\cite{Vanderheyden}. Lattice simulations of the model are also possible
because the fermion determinant is real and non-negative for all chemical
potentials. One hopes that these developments will uncover generic phenomena
that will also apply to QCD at nonzero chemical potential. Furthermore, most
of the theories that aim at a description of the phase diagram of 3-color QCD
at nonzero baryon chemical potential and temperature can be implemented to
describe the phase diagram of 2-color QCD at nonzero baryon chemical potential
and temperature. Therefore, an important motivation to conduct Lattice
simulations is to check the validity of these theories in the 2-color case. If
a theory is not correct in the two-color case, it is very doubtful that the
predictions of the corresponding 3-color theory can be trusted.

Preliminary lattice simulations of the $SU(2)$ model with four species of
quarks, simulation data and an Effective Lagrangian analysis of aspects of
the $T$-$\mu$ phase diagram were recently published \cite{DK2000,PLB}. It is
the purpose of this article to present the continuation of that work on larger
lattices, closer to the theory's continuum limit with a focus on the system's
phase diagram at nonzero $\mu$ and $T$. Very early work on this model at
finite $T$ and $\mu$ was performed by \cite{t+mu}. A simulation study of the
spectroscopy of the light bosonic modes will be presented elsewhere
\cite{inprep}.

In our exploratory study \cite{PLB} of this model's phase diagram, we found
an unanticipated result. At fixed but small quark mass, which insures us that
the pion has a nonzero mass and chiral symmetry is explicitly broken, we found
that at high $T$ and $\mu$ there is a line of transitions separating a
quark-gluon plasma phase  from one with a diquark condensate. Along this line
there is a tricritical point, labeled $2$ in Fig. 1, where the transition
switches from being second order (and well described by mean field theory) at
relatively low $\mu$ to a first order transition at an intermediate $\mu$
value. We will present evidence here for metastability along the line of high
$\mu$ and $T$. We have argued elsewhere \cite{PLB} that this simulation
result, the existence of a tricritical point $2$, has a natural explanation in
the context of chiral Lagrangians. Following the formalism of \cite{Toublan}
we argued that trilinear couplings among the low lying boson fields of the
Lagrangian become more significant as $\mu$ and $T$ increase and they can
cause the transition to become first order at a $\mu$ value in the vicinity of
the results found in the simulation.  Using hindsight, this behavior should not
have come as a surprise. $\mu$ plays the r\^{o}le of a second `temperature' in
this theory in that it is a parameter that controls diquark condensation, but
does not explicitly break the (quark-number) symmetry. The existence of two
competing `temperatures' is a characteristic of systems which exhibit
tricritical behavior. A tricritical point is also found in Chiral Perturbation
Theory, as well as in a new Random Matrix Theory model \cite{STV2,KTV}.

%\enlargethispage*{1000pt}

\begin{figure}[htb!]
\centerline{
\epsfxsize 5 in
\epsfysize 3 in
\epsfbox{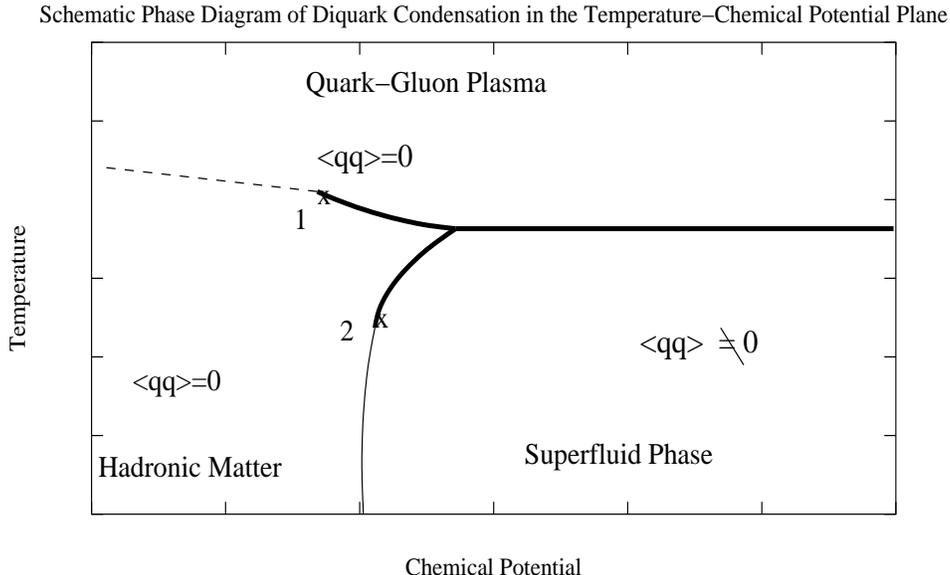}
}
\caption{Schematic Phase Diagram in the $T$-$\mu$ Plane. The thin(thick) line
consists of second(first) order transitions. The dashed line denotes a crossover.
Point 1 labels a critical point and Point 2 labels a tricritical point.
The existence and position of point 1 is
a matter of conjecture in the two color model.
}
\label{fig:phase}
\end{figure}

%\pagebreak

The conjectured phase diagram of Fig.~\ref{fig:phase} also has a critical
point labeled $1$ which connects the line of first order transitions to the
dashed line of crossover phenomena extending to the $\mu=0$ axis where we
expect a finite T chiral transition between conventional hadronic matter and
the quark-gluon plasma. For sufficiently light quarks in the four flavor SU(2)
model, the $\mu=0$ transition is known to be first order both from
theoretical arguments \cite{PW,Wirstam} and simulations \cite{JBK}. For quark
mass $m=0.05$ we find that the transition is smoothed out to a crossover
region as $\mu$ is increased until the dynamics favors diquark condensation.
Clearly, the results will depend on the precise value of the quark mass $m$.
At $m=0.05$ we find that the transition at $\mu=0.10$ is as smooth as that at
$\mu=0.0$, while the transition at $\mu=0.20$ is noticeably sharper. This
result suggests that if $m$ were chosen somewhat smaller than $m=0.05$, the
scenario with the critical point 1 and the first order line might emerge.
However, this critical point 1 might be absent so the crossover line would
intersect the curve separating the diquark condensation phase from
the normal phase. We will be pursuing this suggestion
in separate simulations. The simulations done here are not accurate enough to
probe the details of the region of the phase diagram Fig.~\ref{fig:phase}
where the two first order lines split apart. Simulations at lower quark masses
on larger lattices might be needed.

In QCD at nonzero baryon chemical potential Fodor and Katz \cite{FK} have
estimated the position of the critical point 1 using a modified, multivariable
Glasgow algorithm \cite{IAN}. It will be interesting to find point 1 in the
four flavor SU(2) model at smaller $m$ values using our conventional
simulation methods and then checking whether the Fodor-Katz method can find it
with competitive accuracy. If point 1 is absent, the Fodor-Katz method should
  be able to find the intersection of the crossover line with the
  curve delimiting the diquark phase.

We also present extensive and accurate simulation results on a $16^4$
lattice, essentially at vanishing $T$. The critical indices of the transition
at nonzero $\mu$ are measured and  favor mean-field scaling
in agreement with Chiral Perturbation Theory, analyzed through
one loop corrections, and simulations of 2-color QCD in the strong coupling
limit \cite{aadgg}. The critical indices $\beta_{mag}$ and $\delta$ are
emphasized. Equation of State fits, suggested by Chiral Perturbation Theory,
also support these conclusions. As a third method to determine the
critical indices, we use the static scaling hypothesis. This analysis is also
compatible with mean field, although this approach is unable to predict the
index $\delta$ accurately because of its sensitivity to estimates of the
critical chemical potential, $\mu_c$. Unfortunately, because simulations at
small $\lambda$ are extremely costly in computing requirements, we have not
been able to approach the critical point closely enough to make definitive
measurements, insensitive to the detailed nature of the critical scaling.
In a recent paper in which we study quenched versions of this and closely
related theories, we have indicated how sensitive measurements of critical
scaling are to the form of the scaling variable \cite{KS_iso}. This paper
does indicate, however, that there are classes of scaling variables for
which there exist extended regions outside of the true scaling window, where
the order parameter scales with the correct critical exponents.

\section{Simulation Method and The Algorithm}

The lattice action of the staggered fermion version of this theory is:

\begin{equation}
S_f = \sum_{sites}\left\{\bar{\chi}[D\!\!\!\!/\,(\mu) + m]\chi
+ \frac{1}{2}\lambda[\chi^T\tau_2\chi + \bar{\chi}\tau_2\bar{\chi}^T]\right\}
\end{equation}

\noindent where the chemical potential $\mu$ is introduced by multiplying
links in the $+t$ direction by $e^\mu$ and those in the $-t$ direction by
$e^{-\mu}$ \cite{First}. The diquark source term (Majorana mass term) is added
to allow us to observe spontaneous breakdown of quark-number on a finite
lattice. The parameter $\lambda$ and the usual mass term $m$ control the
amount of explicit symmetry breaking in the lattice action. We will be
particularly interested in small values of $\lambda$ and the extrapolation
$\lambda \rightarrow 0$ for a given $m$ to produce an interesting, realistic
physical situation. This is the system that has been studied analytically using
effective Lagrangians at non-zero chemical
potential $\mu$ \cite{Toublan,SUNY}, and non-zero temperature \cite{STV2}.

Integrating out the fermion fields in Eq.1 gives:

\begin{equation}
pfaffian\left[\begin{array}{cc} \lambda\tau_2     &    {\cal A}       \\
                                     -{\cal A}^T        &    \lambda\tau_2
\end{array}\right] = \sqrt{{\rm det}({\cal A}^\dagger {\cal A} + \lambda^2)}
\end{equation}
where
\begin{equation}
           {\cal A} \equiv  D\!\!\!\!/\,(\mu)+m
\end{equation}

Note that the pfaffian is strictly positive, so that we can use the hybrid
molecular dynamics \cite{HMD} method to simulate this theory using ``noisy''
fermions to take the square root, giving $N_f=4$.

For $\lambda = 0$, $m \ne 0$, $\mu \ne 0$ we expect no spontaneous symmetry
breaking for small $\mu$. For $\mu$ large enough ($\mu > m_\pi/2$ according to
most approaches including Chiral Perturbation Theory
\cite{Toublan,SUNY,STV1}) we expect spontaneous breakdown of quark
number and one Goldstone 
boson -- a scalar diquark. (The reader should consult \cite{Hands}  for a full
discussion of the symmetries of the lattice action, remarks about
spectroscopy, Goldstone as well as pseudo-Goldstone bosons, and for early
simulations of the 8 flavor theory at $\lambda=0$.)

\section{Simulation Results and Analysis}

We now consider the simulation results for the $N_f=4$ theory on $16^4$, $8^4$
and $12^3 \times 6$ lattices, measuring the chiral
($\langle\bar{\chi}\chi\rangle$) and diquark condensates
($\langle\chi^T\tau_2\chi\rangle$), the fermion number density, the
Wilson/Polyakov line, etc.

\begin{figure}[htb!]
\epsfxsize=6in
\centerline{\epsffile{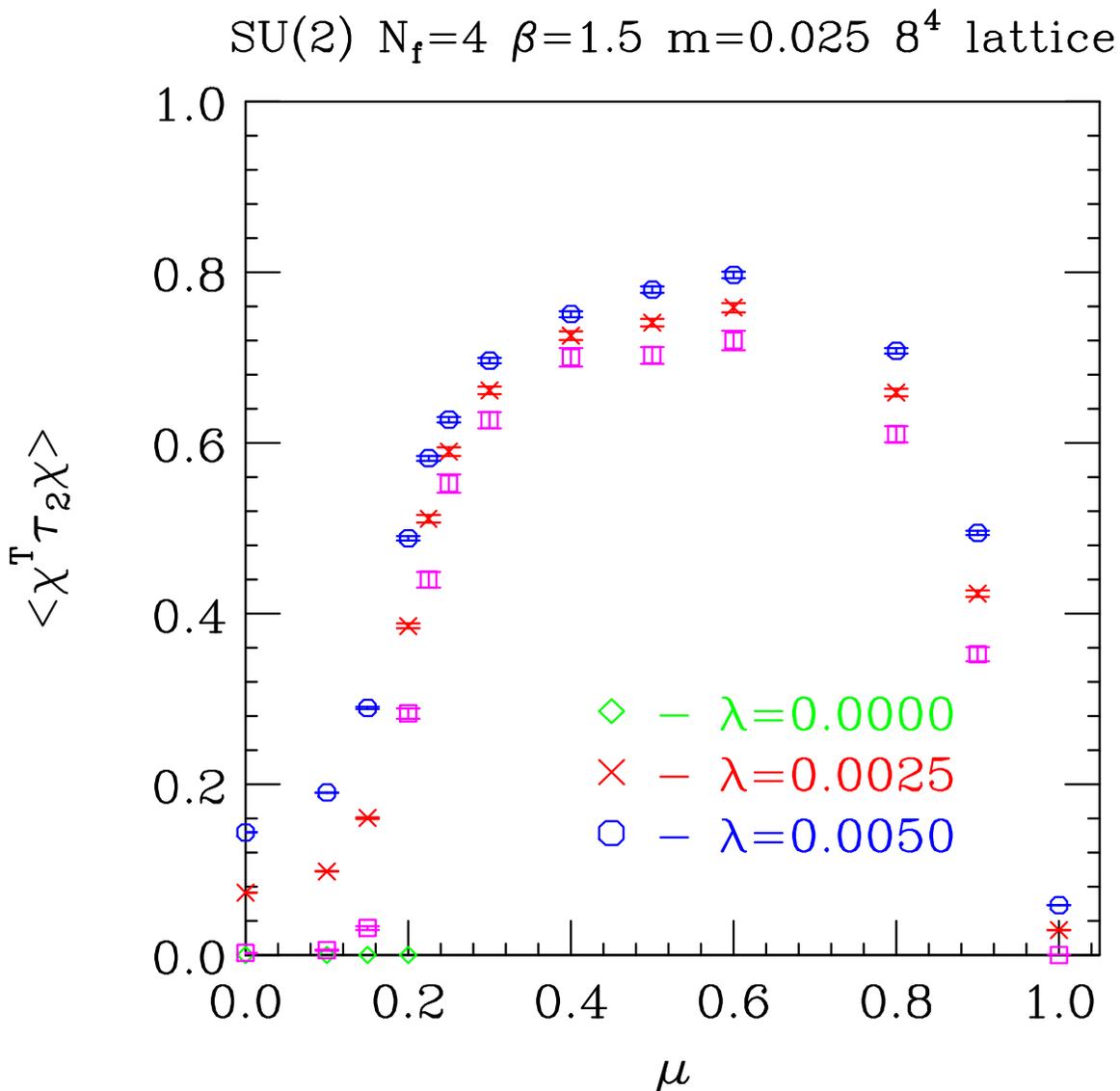}}
\caption{$\langle\chi^T\tau_2\chi\rangle$ as a function of $\mu$ on an $8^4$
lattice at $\beta=1.5$, $m=0.025$. The unlabeled squares are a linear
extrapolation to $\lambda=0$}
\label{fig:pt2p_8}
\end{figure}
Let us first briefly mention our $8^4$ simulations. These were performed at
$\beta=1.5$, a relatively strong coupling, and $m=0.025$. These extend our
earlier work with the same $\beta$ but $m=0.1$. Using a smaller quark mass
increases the validity of chiral perturbation analyses. In addition, we are
performing simulations on a $12^3 \times 24$ lattice with the same parameters
to examine the spectrum of Goldstone and pseudo-Goldstone bosons of this theory.
For this purpose it is necessary to have the pion mass significantly lighter
than that of non-Goldstone, non-pseudo-Goldstone excitations. We have simulated
at $\lambda=0.0025$, and $0.005$. Our measurements of the diquark condensate are
given in figure~\ref{fig:pt2p_8}.
Each `data' point represents a 2000 molecular-dynamics time-unit simulation.
The fact that a simple linear extrapolation of our measurements to $\lambda=0$
is very small, not only at $\mu=0$, where we know that the condensate vanishes,
but also at $\mu=0.1$ leads us to believe that the condensate vanishes in this
limit, probably up to $\mu \sim 0.2$. (Note that is is known that there is some
small but finite value of $\mu$, below which this condensate vanishes
\cite{aadgg}.) For $\mu$ much greater than $0.2$ but less than the saturation
value it is reasonably clear that the condensate does not vanish in the limit
as $\lambda \rightarrow 0$ (on an infinite lattice). Hence there is a phase
transition for some value of $\mu$ to a state where quark number is
spontaneously broken by a diquark condensate. The transition occurs over a
relatively small range of $\mu$, but does not appear to be first order.
Because of the relatively small scaling window this implies, we will not be
able to analyze critical scaling until we perform measurements at more $\mu$
values within this scaling window.

The chiral condensate remains approximately constant up to $\mu=\mu_c$ after
which it falls rapidly approaching zero for large $\mu$. The fermion density
rises slowly and approximately linearly from zero for $\mu \ge \mu_c$. Above
$\mu \approx 0.6$, it increases more rapidly approaching its saturation value
of $2$ at $\mu \approx 1.0$. 

We now turn to measurements on a $16^4$, `zero temperature' lattice. We
simulated the $SU(2)$ model at a relatively weak coupling $\beta=1.85$, within
the theory's apparent scaling window, in order to attempt to make contact with
the theory's continuum limit. The quark mass was $m= 0.05$, as in \cite{PLB},
and a series of simulations were performed at $\lambda = 0.0025$, $0.005$, and
$0.01$ so that our results could be extrapolated to vanishing diquark source,
$\lambda=0$. Simple linear and spline extrapolations were used and these
procedures appeared to be sensible away from the transition. None of the
conclusions to be drawn here will depend strongly on the limit. In fact,
everything we say could be gathered from our data at a fixed (small) $\lambda$
value, $0.005$ or $0.0025$, say. However, only in the limit of vanishing
$\lambda$ do we expect real diquark phase transitions (at least when the
transitions are second order), so it is particularly interesting and relevant
to investigate the  $\lambda \rightarrow 0$ limit. The quantitative agreement
between fits at fixed, small $\lambda$ values and those using extrapolated
data enforce the prevalent idea that simple extrapolation procedures are
reliable except in the immediate vicinity of the critical point
where non-analyticities are inevitable.

Because we are forced by practical considerations to work at $\lambda$ values
large enough that the condensate is relatively smooth over the transition
region, we have to deduce critical scaling from the curvature of the
condensate further from $\mu_c$ than we would like. Thus we run the risk that
we are outside the scaling window and the apparent scaling we are seeing does
not represent true critical scaling. However, we saw for quenched theories
\cite{KS_iso} at relatively weak couplings, that the form of the natural
scaling variable is such that it has an inflection point when expressed in
terms of the simplest scaling variable $\mu-\mu_c$. This leads to an extended
domain where the order parameter scales with the true critical exponents. It
is reasonable to assume that dynamical theories behave similarly.

In Table I we present the raw data on the $16^4$ lattice at the three
$\lambda$ values. Table II exhibits the results of extrapolating this data
to $\lambda=0$. Typically, one thousand time units of the Hybrid
Molecular Dynamics algorithm were accumulated to generate these data sets. The
error bars in the data sets of Table I account for the correlations in the raw
data sets. The low $\lambda$ runs were our most CPU intensive.

Since the Hybrid algorithm suffers from systematic errors proportional to the
square of the discrete time step used in integrating its stochastic
differential equations forward in Monte Carlo time \cite{HMD}, we were forced
to a small discrete time step. The simulations reported here used time steps
as small as $dt=0.0025$ to control these errors. Even then, because our
statistical errors are so small, these systematic errors might not be 
negligible. We will later indicate cases where we used $dt$ values which were
larger than optimal and have seen signs of such errors.

\begin{figure}[htb!]
\centerline{
\epsfxsize 5 in
\epsfysize 3 in
\epsfbox{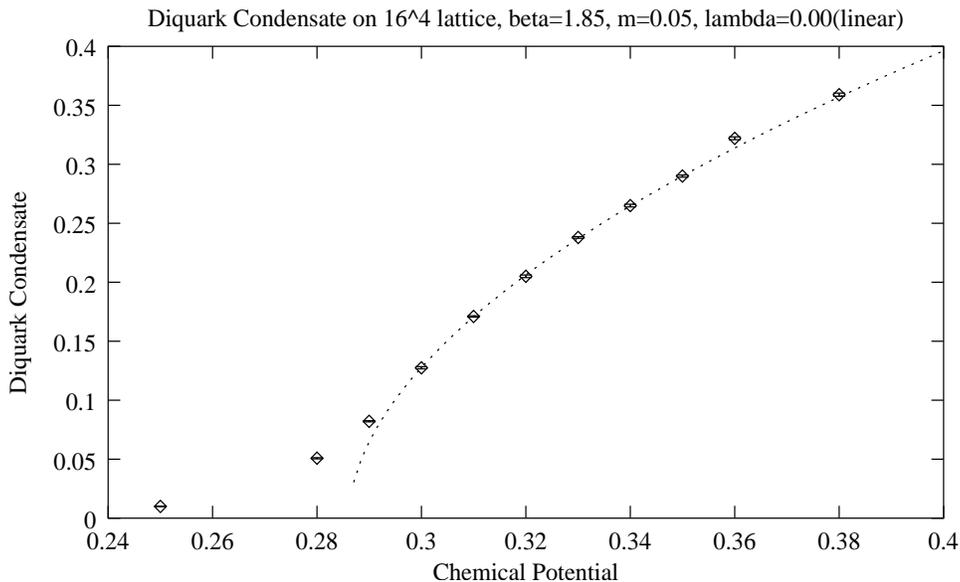}
}
\caption{Diquark Condensate vs. $\mu$}
\label{fig:linear}
\end{figure}

Now let us discuss some results. In Fig.~\ref{fig:linear} we show the diquark
condensate linearly extrapolated to $\lambda=0$ plotted against the
chemical potential $\mu$. The quark mass $m$ was fixed at $0.05$, the coupling
was $\beta=1.85$ and the linear extrapolation used the raw data at
$\lambda=0.0025$ and $0.005$.

\noindent We see evidence to a quark-number violating second order phase
transition in this figure. The dashed line is a power law fit from $\mu=0.30$
to $0.35$ which predicts the critical chemical potential of $\mu_c =
0.2860(2)$. The power law fit is good, its confidence level is $48$ percent,
and its critical index is $\beta_{mag}= 0.54(3)$ which is consistent
with the mean field result $\beta_{mag}=1/2$, predicted by chiral perturbation
theory \cite{Toublan}, \cite {SUNY} including one loop corrections. Note that
the quark mass is fixed at $m=0.05$ throughout this simulation, so chiral
symmetry is explicitly broken and the transition of Fig. 2 is due to quark
number breaking alone.

\begin{figure}[htb!]
\centerline{
\epsfxsize 5 in
\epsfysize 3 in
\epsfbox{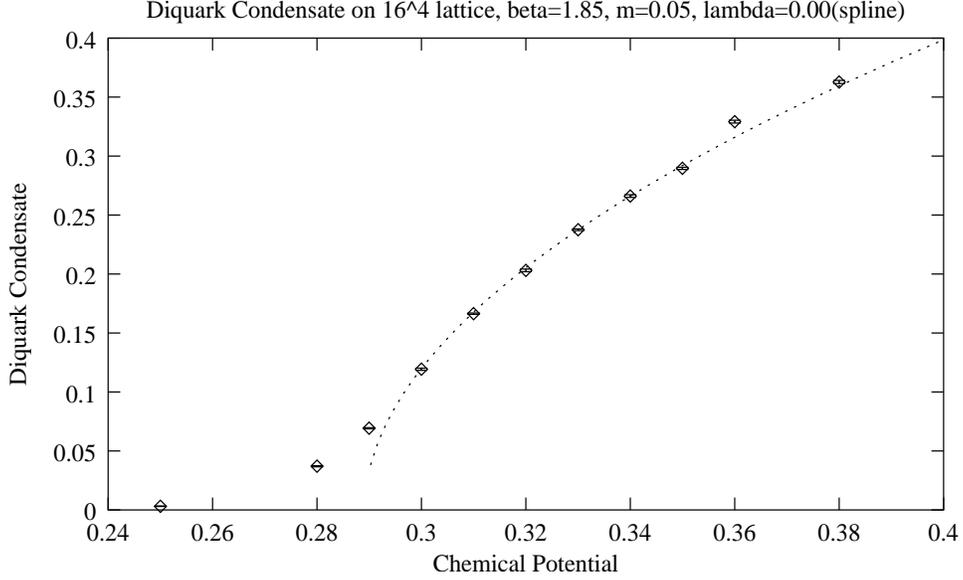}
}
\caption[]{Diquark Condensate vs. $\mu$}
\label{fig:spline}
\end{figure}

In figure~\ref{fig:spline} we show the diquark data extrapolated to
$\lambda=0$ quadratically using 
the raw data at $\lambda=0.010$, $0.005$ and $0.0025$. The fit to the
data at $\mu$ ranging 
from $0.30$ through $0.35$ is also shown. The fit
predicts a critical point $\mu_c=0.2870(2)$, has a confidence level of $47$
percent and has a critical exponent of $\beta_{mag}=0.54(4)$, again in 
agreement with mean field theory.

\begin{figure}[htb!]
\centerline{
\epsfxsize 5 in
\epsfysize 3 in
\epsfbox{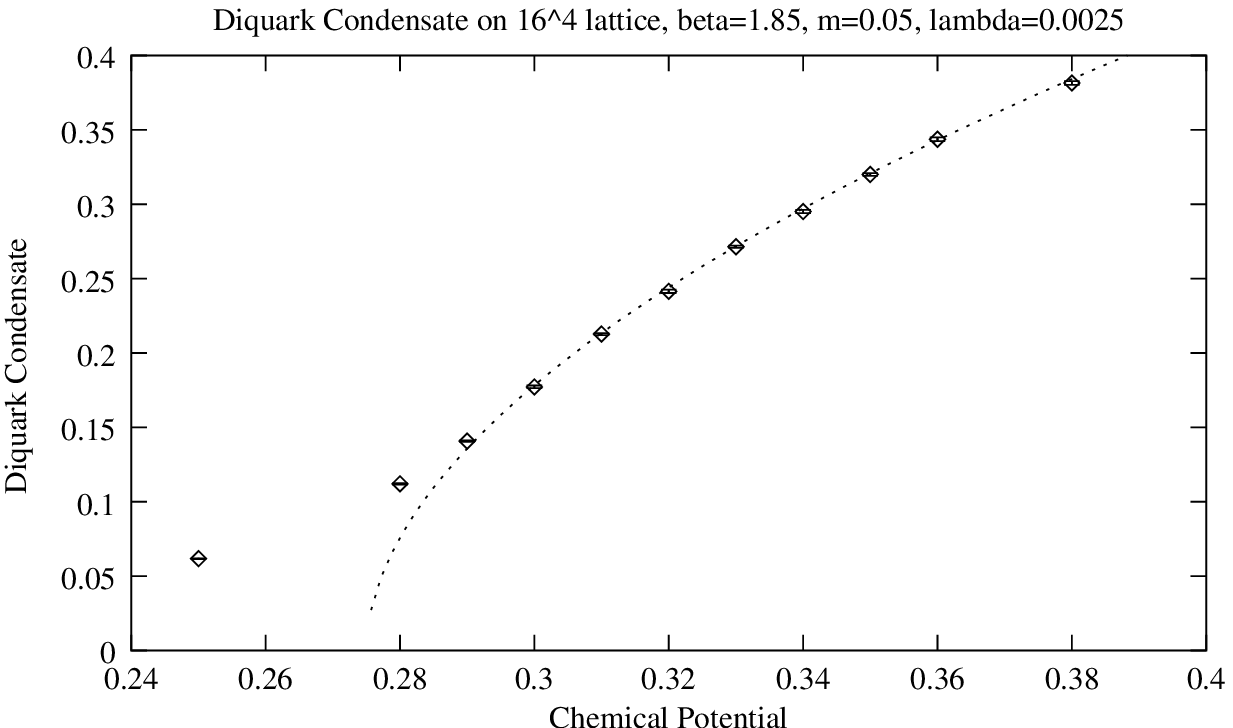}
}
\caption{Diquark Condensate vs. $\mu$.}
\label{fig:l0025}
\end{figure}

Further evidence for $\beta_{mag}=1/2$ can be obtained directly from the raw
data without any extrapolations. For example, in figure~\ref{fig:l0025} we
show the $\lambda=0.0025$ data and a power law fit ranging form $\mu=0.30$
through $0.38$.
We find $\mu_c=0.2749(2)$, $\beta_{mag}=0.54(5)$ with a
confidence level of $24$ percent.
We also considered extrapolations of the form 
$\langle\chi^T\tau_2\chi\rangle = \rho$ where $\rho$ is the solution of an
equation of the form $\rho^3 = a \: \rho + b \: \lambda$ with $a$ and $b$
determined as functions of $\mu$ by performing an independent fit at each
$\mu$. This form is suggested my mean-field scaling forms, which would predict
that $a$ should be proportional to $\mu-\mu_c$ close to the transition.
However, these fits were so poor that this approach was abandoned.

\begin{figure}[htb!]
\centerline{
\epsfxsize 5 in
\epsfysize 3 in
\epsfbox{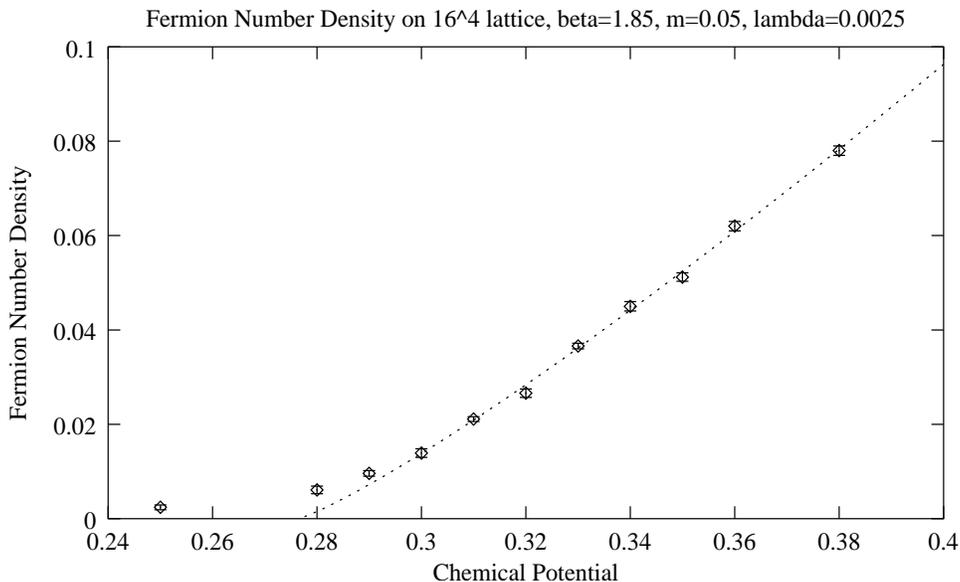}
}
\caption{Fermion Number Density vs. $\mu$.}\label{fig:j0_0025}
\end{figure}

In addition, we considered the fermion number density which is
predicted to rise linearly from $\mu_c$ 
and found good agreement with this expectation, as shown in 
figure~\ref{fig:j0_0025}.

The fit predicts the critical point $\mu_c = 0.2765(5)$, with a critical index
$1.17(7)$ and a confidence level of $13.1$ percent. Plots and fits of the
fermion number density at the other $\lambda$ values were equally good. In fact
the fermion number density shows relatively little $\lambda$
dependence. This is in agreement with Chiral Perturbation Theory
\cite{SUNY}.

\begin{figure}[htb!]
\centerline{
\epsfxsize 5 in
\epsfysize 3 in
\epsfbox{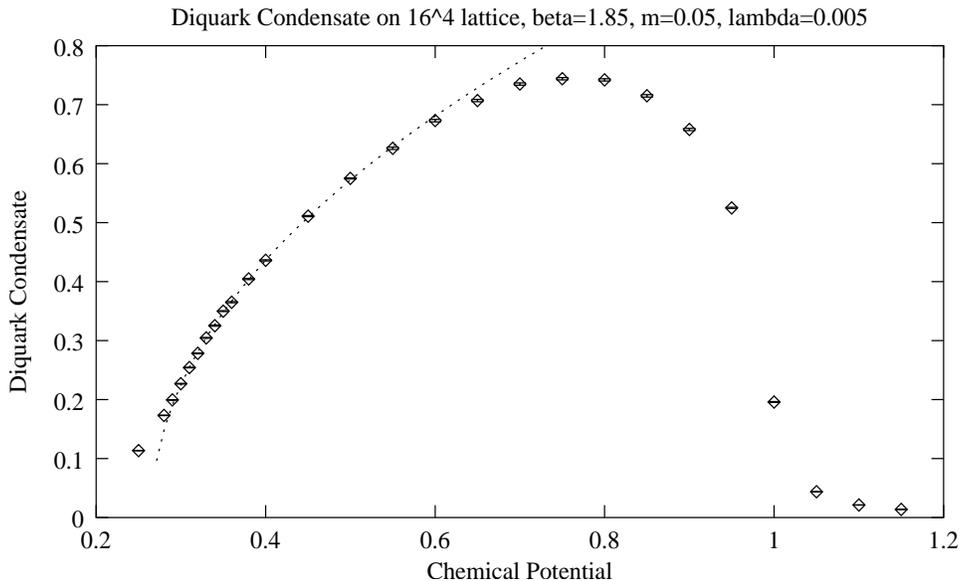}
}
\caption{Diquark Condensate vs. $\mu$.}\label{fig:l005}
\end{figure}

Finally, we considered the $\lambda=0.005$ simulations and extended
the measurements up to large 
$\mu \rightarrow 1.2$, shown in figure~\ref{fig:l005}. We see the curious
phenomenon observed in the past: as 
$\mu$ becomes large, the condensate falls to zero. The point is that once 
$\mu$ becomes
of order unity, the discreteness of the lattice comes into play and
suppresses the density of states
which forces the condensate to fall toward zero. The quark-number density
approaches 2 (per lattice site), the maximum allowed by Fermi statistics.
This is a lattice artifact.
Nonetheless, we can
try a power law fit to the diquark condensate 
over the entire region of $\mu$ from $0.30$ to $0.60$ where the curve
is increasing. The result is shown in the figure 
and the fit predicts $\beta_{mag}=0.49(7)$ and $\mu_c=0.2651(2)$. Since 
$\chi^2/dof \sim 5$, the fit is poor, but the deviations from this fit are
small enough that this fit should be considered a reasonable zeroth order
approximation.

In summary, $\beta_{mag}$ measurements are relatively easy and
self-consistent. For all the parameters 
and procedures we tried, we found good agreement with mean field
theory. It was important, however, to be working 
at relatively weak coupling, $\beta=1.85$, in the theory's scaling
window where such an extended scaling window exists.
Experience has shown that at stronger coupling, the true scaling is described
by effective Lagrangians closer to the non-linear sigma model form, where
the scaling window in $\mu-\mu_c$ is small and the extended scaling regime
exhibits scaling in $\mu-\mu_c$ but with an apparent critical exponent 
$\beta_m$ which is half the true value \cite{aadgg,KS_iso,KS_iso2}.

A second critical index in which we are interested is $\delta$, which controls
the response of the system to symmetry breaking fields at the critical point
$\mu_c$,

\begin{equation}
\langle\chi^T\tau_2\chi\rangle=A \lambda^{1/\delta}
\end{equation}

\noindent in the limit of small $\lambda$. Mean field theory predicts
$\delta=3$. This scaling law Eq. 4 is usually difficult to use in practice
because first it requires accurate estimates of $\mu_c$, and second it
generally suffers from a small scaling window in $\lambda$. Not having a
precise estimate of $\mu_c$ is probably our biggest problem, since as we have
shown in \cite{KS_iso}, fits to the extended scaling regime can underestimate
$\mu_c$ by $\approx 8\%$. In order to obtain some useful data, however, we
simulated the $16^4$ lattice for four estimates of $\mu_c$ each over
a range of $\lambda$ from $0.002$ to $0.010$. First we note that the chosen
$dt$ for the $\lambda=0.002$ simulations was too large and although plotted
on our graphs, these points were excluded from our fits.

We considered $0.2855$, $0.2870$, $0.2920$ and $0.2970$ as estimates of
$\mu_c$ and tried fits of the form $A \lambda^{1/\delta} + B$. The data and
fits are shown in the next four figures, which we discuss in turn.

\begin{figure}[htb!]
\centerline{
\epsfxsize 5 in
\epsfysize 3 in
\epsfbox{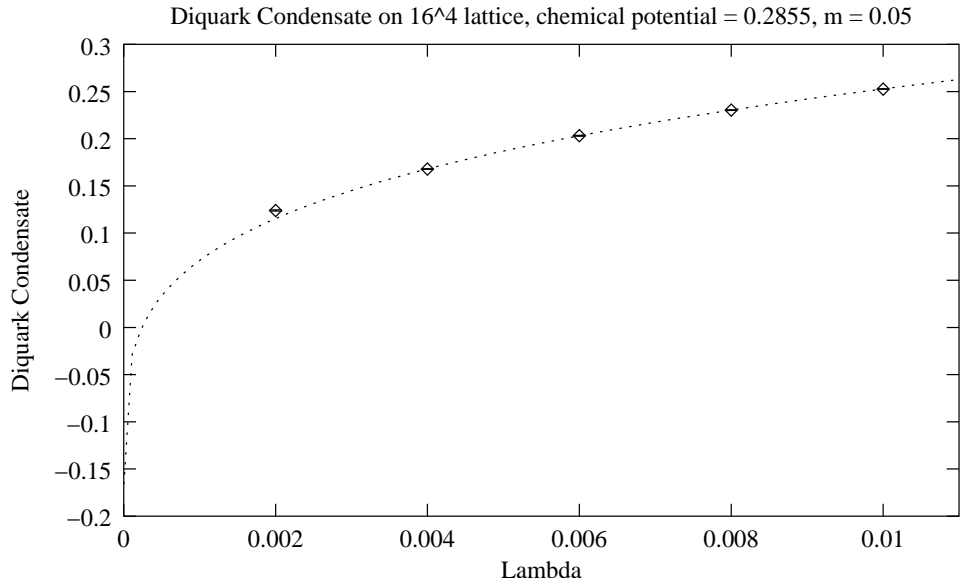}
}
\caption[]{Diquark Condensate vs. $\lambda$ for $\mu_c=0.2855$}
\label{fig:2855}
\end{figure}

The $\mu=0.2855$ data predicted $\delta=4.0(6)$ with a confidence
level of $91$ percent, but with a large, negative 
constant $B=-0.17(7)$. The fit considers the data from $\lambda$
ranging from $0.004$ to $0.010$ because the data 
at $\lambda=0.002$ has large $dt^2$ errors. These results are shown in
figure~\ref{fig:2855} 
%suffered from finite size effects, as we shall see
%below. 
%One might have been suspicious 
%of the $\lambda=0.002$ data because it is so close to the critical
%point that its correlation length is 
%greater than $8$ lattice spacings and is inappropriate for a $16^4$
%simulation. 
The fact that $B$ is large and 
negative suggests that this estimate for $\mu_c$ is too low.

\begin{figure}[htb!]
\centerline{
\epsfxsize 5 in
\epsfysize 3 in
\epsfbox{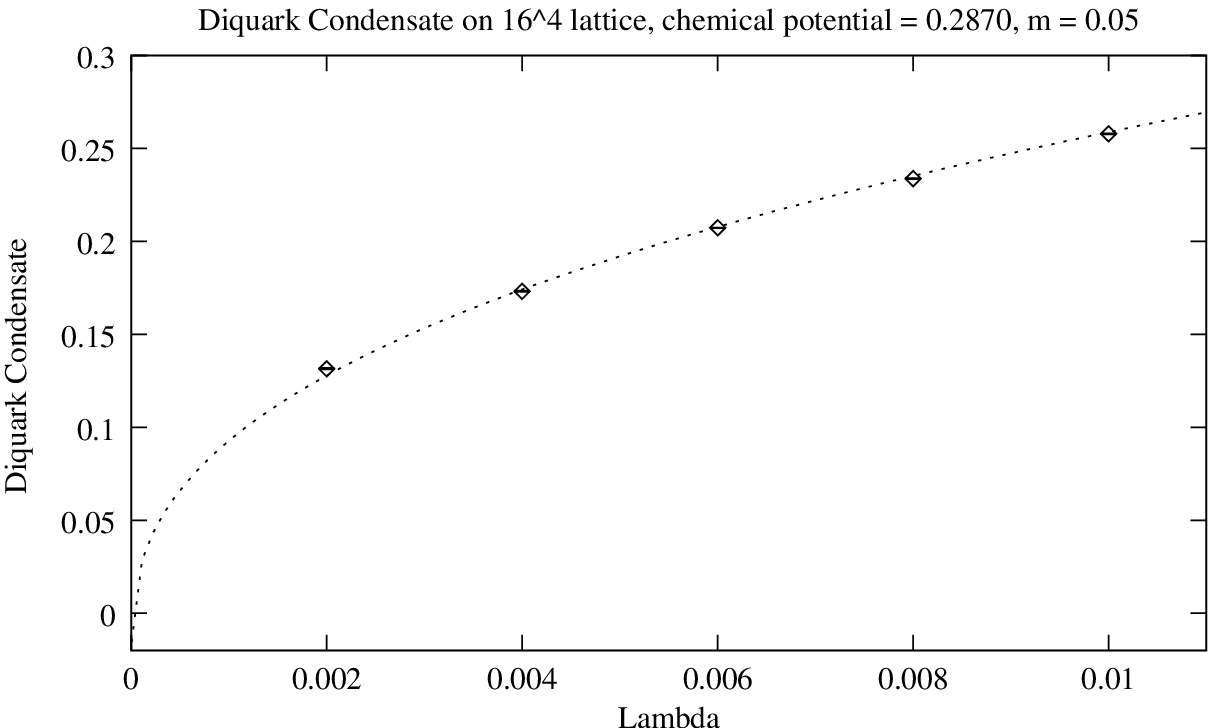}
}
\caption[]{Diquark Condensate vs. $\lambda$ for $\mu_c=0.2870$}
\label{fig:2870}
\end{figure}

\begin{figure}[htb!]
\centerline{
\epsfxsize 5 in
\epsfysize 3 in
\epsfbox{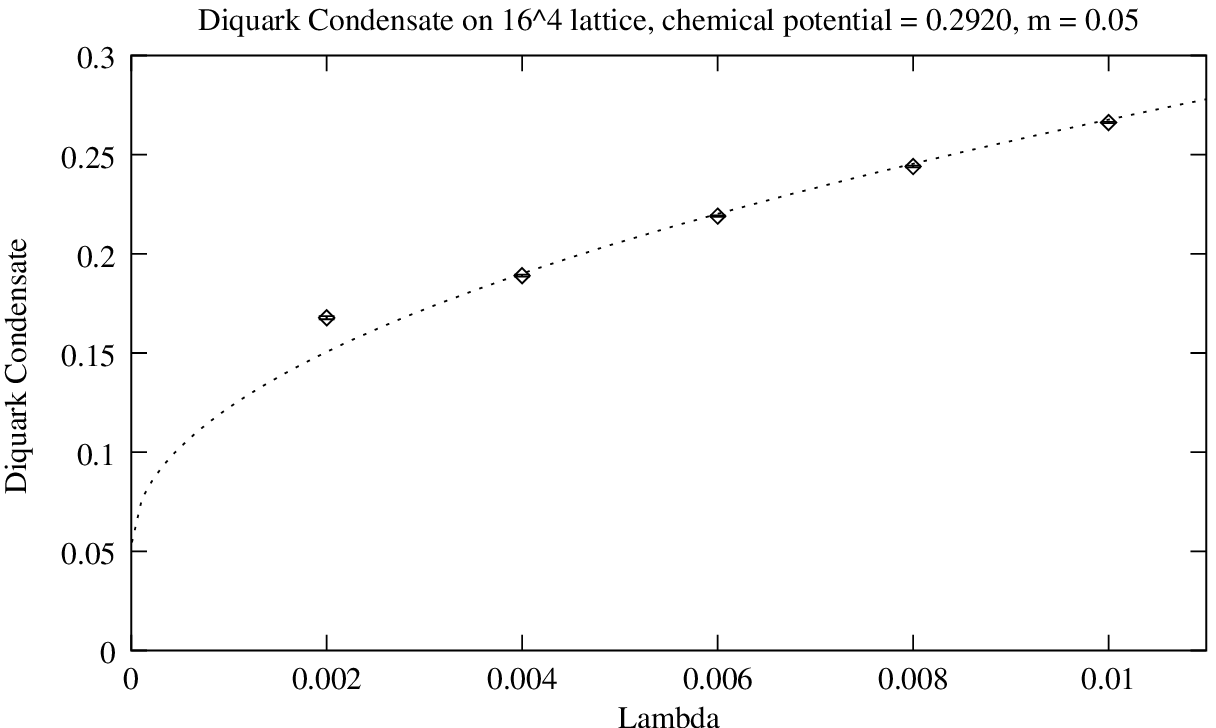}
}
\caption[]{Diquark Condensate vs. $\lambda$ for $\mu_c=0.2920$}
\label{fig:2920}
\end{figure}

\begin{figure}[htb!]
\centerline{
\epsfxsize 5 in
\epsfysize 3 in
\epsfbox{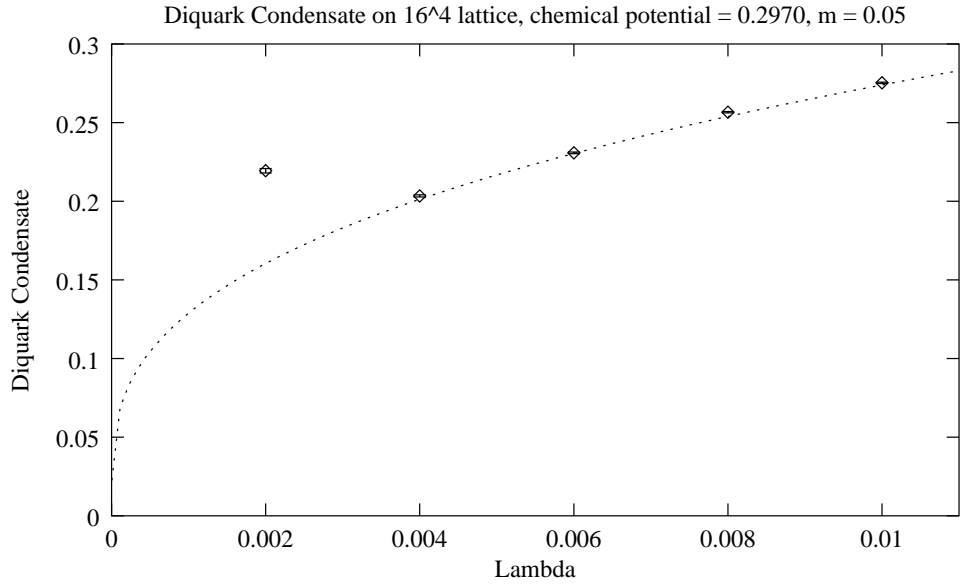}
}
\caption[]{Diquark Condensate vs. $\lambda$ for $\mu_c=0.2970$}
\label{fig:2970}
\end{figure}

In figure~\ref{fig:2870} we consider the same ideas for the estimate
$\mu_c=0.2870$ and now find $\delta=2.5(2)$ 
and $B=-0.02(2)$, but with a rather poor confidence level of a few
percent. For $\mu_c=0.2920$ (figure~\ref{fig:2920}), we find 
$\delta=2.1(2)$, but $B$ is distinctly different from zero,
$B=0.05(2)$. The confidence level is, however, 
very good, $96$ percent. Finally, for $\mu_c=0.2970$ we find
$\delta=2.7(6)$ and $B=0.02(4)$ with a confidence 
level of about one percent. See figure~\ref{fig:2970}.

In summary, $\delta$ is in the mean field `ballpark', but the uncertainly in
the simulation results is large. The results give us no reason to doubt the
application of mean field theory because they are all in the vicinity of
$\delta=3$ with some fits higher and some lower. The sensitivity of the
simulation estimates to $\mu_c$ is considerable. We learn from these
measurements that only a focused set of measurements that determine $\mu_c$
very well will produce a useful estimate of $\delta$.

\section{Comparison with Chiral Perturbation Theory}

It is interesting to compare the lattice data to the theoretical predictions
of Chiral Perturbation Theory.
%Within this framework we will
%analyze the data assuming static scaling and will study the extrapolation of 
%the diquark condensate to zero diquark source, $\lambda \rightarrow 0$.
Two-color QCD at non-zero baryon/quark-number chemical potential has been
thoroughly studied at the tree level and at next-to-leading order within Chiral
Perturbation Theory \cite{Toublan,SUNY,STV1}. At next-to-leading order, it was
found that the phase transition between the normal phase and the superfluid
diquark phase is second order, that the critical chemical potential is given
by $\mu_c=m_\pi/2$ and that the critical exponents %at next-to-leading order
are given by mean-field theory. Furthermore, based on the structure of the
loop corrections, it was conjectured that the critical exponents are given by
mean field at any (finite) order in Chiral Perturbation Theory \cite{STV1}.

Overall, the next-to-leading-order corrections do not dramatically change the
leading-order results \cite{STV1}. The corrections turn out to be
small, as long  as $\mu$ is close to $\mu_c$.
Therefore it should be enough to compare the lattice data to Chiral
Perturbation Theory at leading-order, at least for the lattices sizes,
couplings and parameters considered here.

In Chiral Perturbation Theory at leading order, the diquark condensate at
non-zero diquark source is given by
\begin{equation}
\langle \chi^T \tau_2 \chi \rangle=\langle \bar{\chi} \chi \rangle_0 
\sin \alpha,
\label{diq}
\end{equation}
where $\langle \bar{\chi} \chi \rangle_0$ is the chiral condensate at zero
chemical potential, zero temperature, and zero diquark source, and $\alpha$ is
given implicitly by
\begin{equation}
4 \mu^2 \cos \alpha \sin \alpha = m_\pi^2 \sin(\alpha-\phi).
\label{SadPt}
\end{equation}
In the equation above  $\tan \phi=\lambda/m$, and $m_\pi$ is the mass
of the Goldstone modes at chemical potential $\mu=0$, temperature
$T=0$, and diquark source $\lambda = 0$ \cite{SUNY}. Furthermore the
quark-antiquark (chiral) condensate is given by
\begin{equation}
\langle \bar{\chi} \chi \rangle=\langle \bar{\chi} \chi \rangle_0 \cos \alpha.
\label{qbarq}
\end{equation}

\begin{center}
\begin{figure}[htb!]
\vspace{1cm}
\hspace*{-.1cm}
\epsfig{file=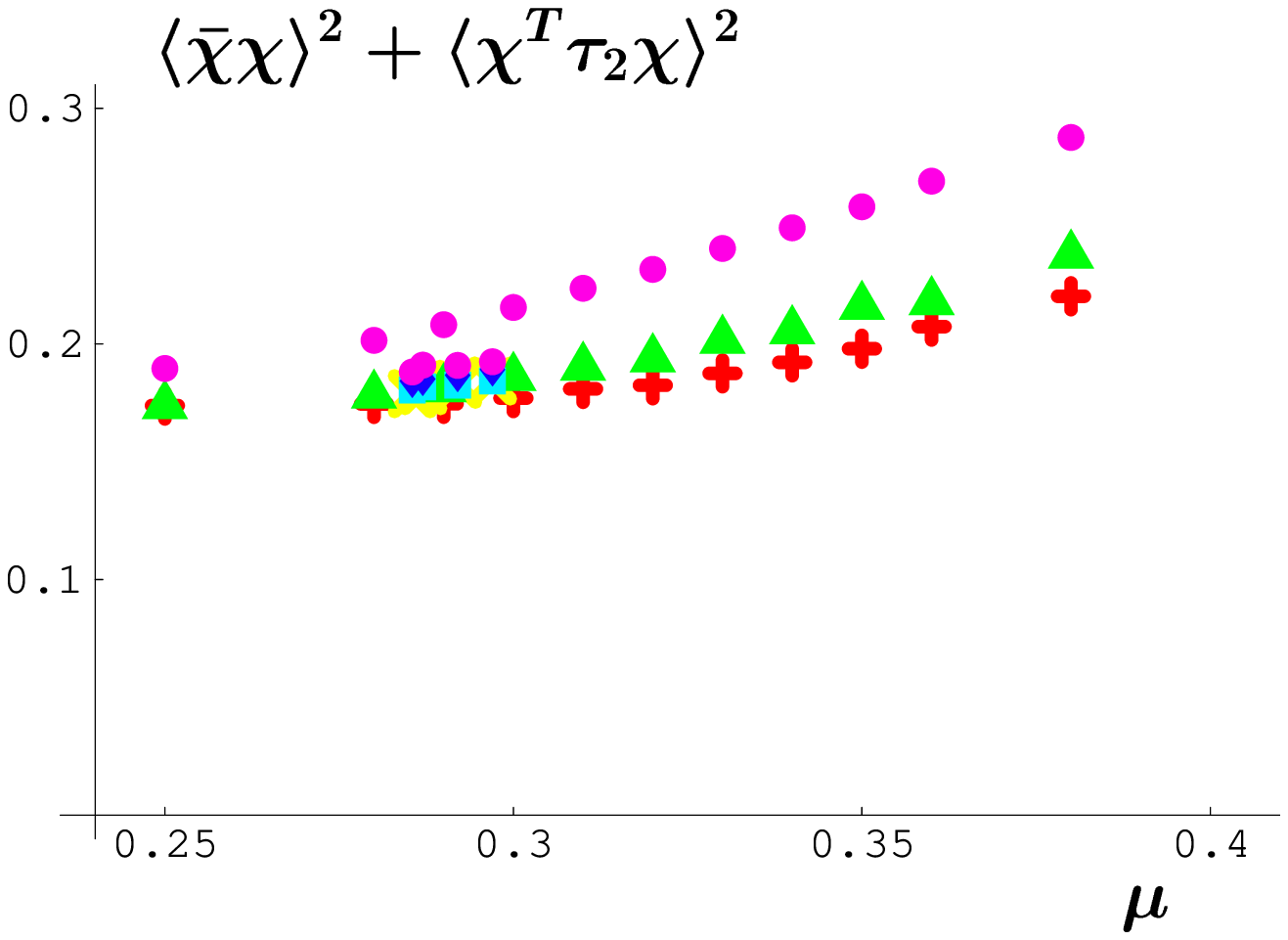, width=13.3cm, height=8cm}\vspace{2cm}
\epsfig{file=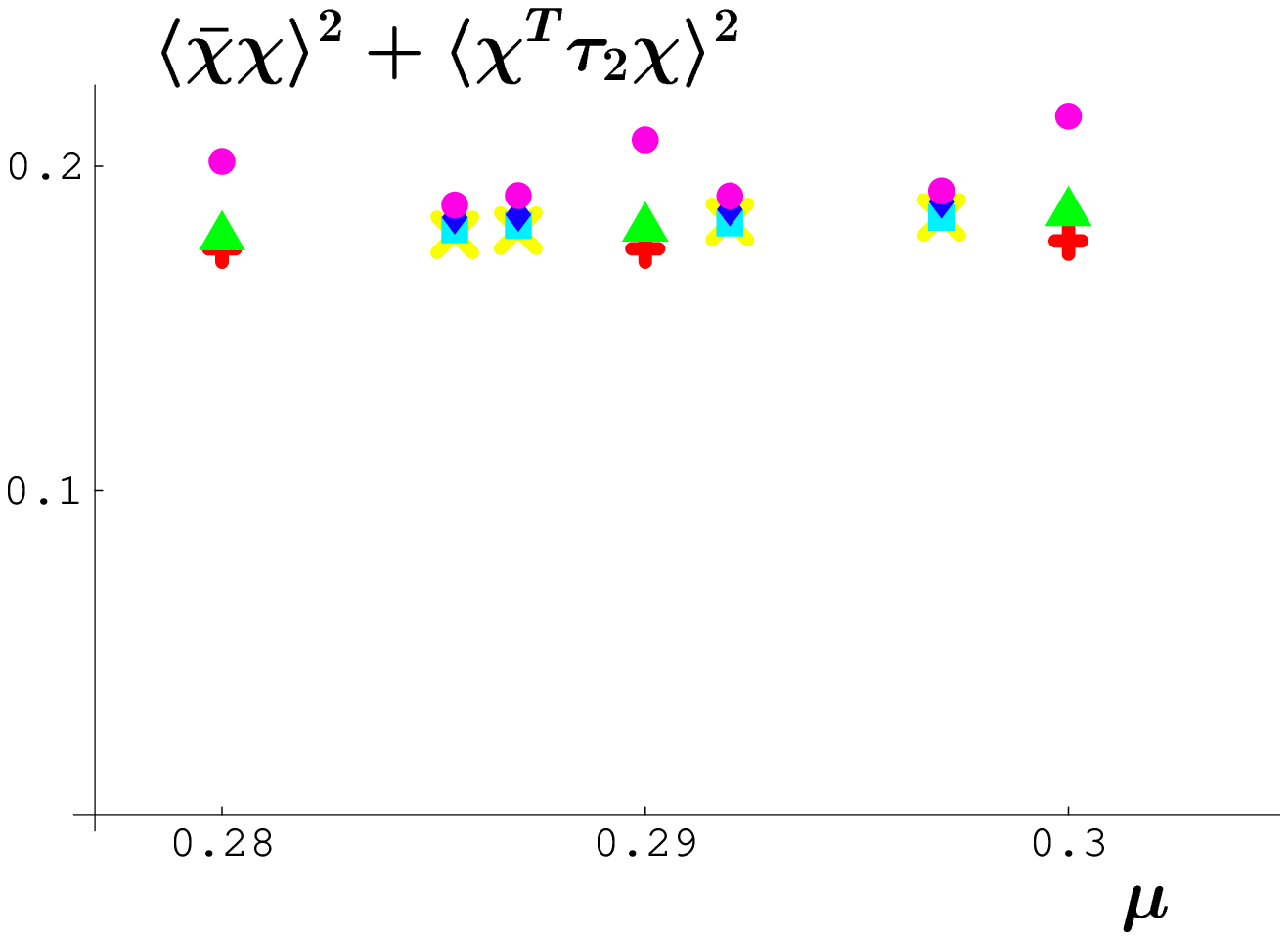, width=13.3cm, height=8cm}
\caption[]{\small $\langle \bar{\chi} \chi \rangle^2+ \langle \chi^T \tau_2
  \chi\rangle^2$ for different diquark
  sources and chemical potentials. The pluses are the 
lattice data for $\lambda=0.025$, the xs for $\lambda=0.004$,
the triangles for $\lambda=0.005$, the squares for
$\lambda=0.006$,  the diamonds for $\lambda=0.008$, and the
dots for $\lambda=0.010$. a) shows that $\langle \bar{\chi} \chi
\rangle^2+\langle \chi^T \tau_2  \chi\rangle^2$ is almost constant for
  $0.28\leq\mu\leq0.30$ and for a given diquark source. b) shows the
  transition region on an expanded scale.}   
\label{fig:norm}
\end{figure}
\end{center}

At nonzero diquark source there is a crossover between the normal
phase and the diquark condensation phase. However, Chiral Perturbation
Theory at leading order predicts that 
$\langle \bar{\chi} \chi \rangle^2+\langle \chi^T \tau_2 \chi \rangle^2$
does not depend on the chemical potential, the diquark source, or the quark 
mass. This relation does not hold at next-to-leading order in Chiral
Perturbation Theory \cite{STV1}. However, for small chemical potential, quark
mass and diquark source, Chiral Perturbation Theory at leading order should give
an accurate description of these observables. Therefore, we expect that 
$\langle \bar{\chi} \chi \rangle^2+\langle \chi^T \tau_2 \chi \rangle^2$
depends slightly on the quark mass and the diquark source, in the same way as
the quark-antiquark condensate depends on the quark mass in three-color QCD at
zero temperature and chemical potential \cite{GL}. The behavior of 
$\langle \bar{\chi} \chi \rangle^2+\langle \chi^T \tau_2 \chi \rangle^2$ is
presented in Figure~\ref{fig:norm}. These graphs confirm what was expected from
Chiral Perturbation Theory: $\langle \bar{\chi} \chi \rangle^2+\langle
\chi^T \tau_2 \chi \rangle^2$ is close to constant for small enough $\mu$ and
$\lambda$. Notice that  it is impossible to see the critical
chemical potential, which is around $0.29$ as determined in the previous
section, by looking at $\langle \bar{\chi} \chi \rangle^2+\langle \chi^T \tau_2
\chi \rangle^2$. The points at $\lambda=0.010$ show the effects of finite
$dt^2$ errors. The points which lie higher were generated in simulations
with $dt=0.01$ while the 4 lower points were generated in simulations with
$dt=0.005$. 

We could compare the diquark condensate obtained on the lattice with the
leading-order result (\ref{diq}). But we notice that within Chiral
Perturbation Theory at leading order we have that
\begin{eqnarray}
\frac{\langle \chi^T \tau_2 \chi \rangle}{\langle \bar{\chi} \chi \rangle}=
\tan \alpha.
\label{ratio}
\end{eqnarray}
This observable can also be used as an order parameter of the diquark
condensation phase. In Chiral Perturbation Theory at leading order, for a given
quark mass, diquark source and chemical potential, the ratio (\ref{ratio}) 
only depends on the value of the pion mass $m_\pi$ through the minimization
equation (\ref{SadPt}). It is therefore a very suitable observable to use in
the comparison between the lattice results and Chiral Perturbation Theory.

\begin{center}
\begin{figure}[htb!]
\vspace{-.7cm} 
\hspace*{-.1cm}
\epsfig{file=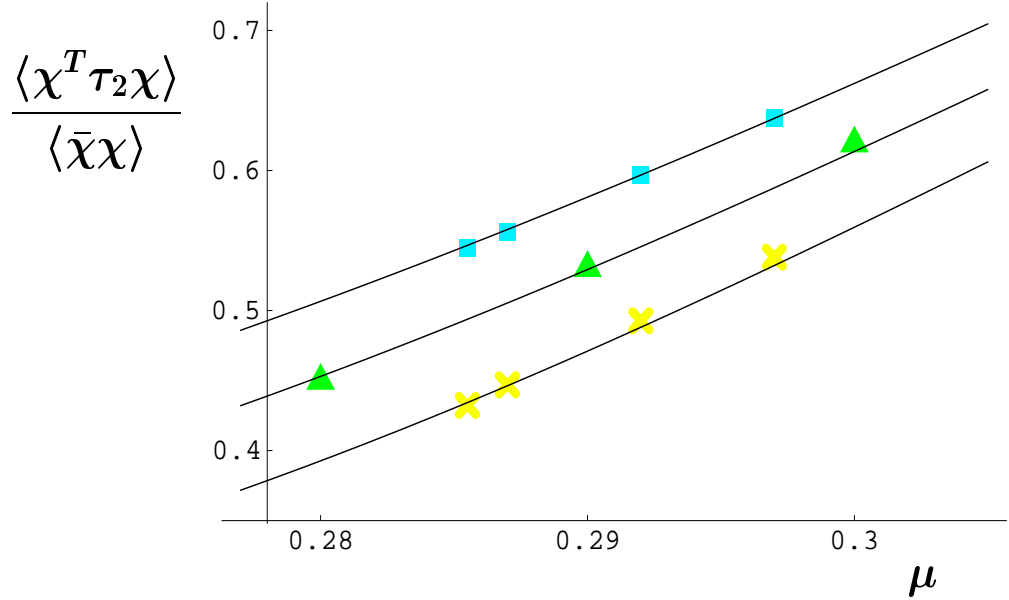, width=13.3cm, height=8cm}\vspace{2cm}
\epsfig{file=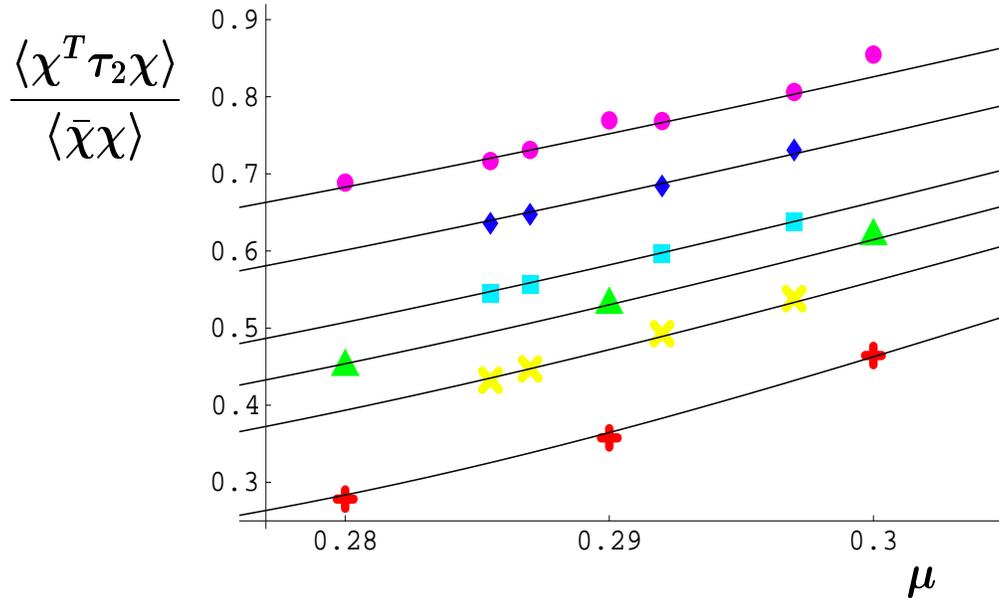, width=13.3cm, height=8cm}
\caption[]{\small Comparison between Chiral Perturbation Theory and
the lattice results 
for  $\langle \chi^T \tau_2 \chi\rangle/\langle \bar{\chi} \chi
\rangle$. The pluses are the 
lattice data for $\lambda=0.025$, the xs for $\lambda=0.004$,
the triangles for $\lambda=0.005$, the squares for
$\lambda=0.006$,  the  diamonds for $\lambda=0.008$, and the dots are for 
$\lambda=0.010$. The results of two different fits with Chiral
Perturbation Theory and the data we used for these fits are depicted
in the solid curves for the different 
diquark sources: a) for $\lambda=0.004,0.005,0.006$ only; b) for all 
$\lambda$s.}
\label{fig:chiral}
\end{figure}
\end{center}

Hence we fit the lattice result for $\langle \chi^T \tau_2 \chi
\rangle/\langle  \bar{\chi} \chi \rangle$ 
with Chiral Perturbation Theory (\ref{ratio},\ref{SadPt}). This is a
{\it one-parameter} fit. Since we use Chiral
Perturbation Theory at leading order to analyze our results, we can only
take into account data corresponding to chemical 
potentials and diquark sources that are small enough so that $\langle
\bar{\chi} \chi \rangle^2+\langle \chi^T \tau_2 \chi \rangle^2$ is
constant for a given diquark source. The previous 
figure shows that we have to limit our analysis to data corresponding
to  $0.28\leq\mu\leq0.30$, where 
$\langle \bar{\chi} \chi \rangle^2+\langle \chi^T \tau_2 \chi \rangle^2$
is close to constant for a given diquark source. If 
we use the 11 points that correspond to $\lambda=0.004$, $0.005$, and
$0.006$, and $0.28\leq\mu\leq0.30$, we get that
$\mu_c=0.3027(1)$. This is an acceptable fit since
$\chi^2/dof=2.3$ (remembering that the norm of the condensate is only
approximately constant over this range.) 
This one-parameter fit and the data we used to perform it are
shown in Figure~\ref{fig:chiral}a.
If we now use the 18 points that correspond to $\lambda=0.0025$, $0.004$,
$0.005$, $0.006$, and
$0.008$, and $0.28\leq\mu\leq0.30$, we get that
$\mu_c=0.3028(1)$, and the fit has a $\chi^2/dof=3.1$.
If we use the 25 points that correspond to $\lambda=0.0025$, $0.004$,
$0.005$, $0.006$, $0.008$, and
$0.010$, and $0.28\leq\mu\leq0.30$, we get that
$\mu_c=0.3026(2)$. The fit is worse than the previous ones with a
$\chi^2/dof=12$. This one-parameter fit and the data we used to perform it are
shown in Figure~\ref{fig:chiral}a. Again we note that the 3 points at
$\lambda=0.010$ which lie above the curve are those with $dt=0.01$ which have
larger $dt^2$ errors, and are largely responsible for this large $\chi^2$.
%I REMOVED THIS BECAUSE PEOPLE WILL SAY -- SO WHAT?!
%Finally if we fit the four points that
%correspond to $\lambda=0.006$, we get an excellent fit with a
%$\chi^2/dof=0.6$ that gives $\mu_c=0.3029(1)$.

Noting that the new estimates for $\mu_c$ are significantly higher than our
previous estimates, we consider the inclusion of the `data' at $\mu=0.31$
and $\mu=0.32$ in our fits. If we include the measurements from $\mu=0.31$
to the 25 points that correspond to $\lambda=0.0025$, $0.004$, $0.005$,
$0.006$, $0.008$ used in the fits of the previous paragraph, the $\chi^2/dof$ 
increases from $3.1$ to $4.9$. Including both the $\mu=0.31$ and $\mu=0.32$
measurements further increases the $\chi^2/dof$ to $5.9$. In both cases the
new $\mu_c$ is entirely consistent with the estimates of the previous paragraph.

In summary, we find that Chiral Perturbation Theory at leading
order describes the data well, with a critical chemical potential given by
$\mu_c=0.3027(3)$. This is somewhat higher than our earlier estimates
in the previous section, and might explain why our $\delta$ estimates
were less than compelling. 
It is also understandable in terms of what was learned from the quenched
studies performed in \cite{KS_iso}. In Chiral Perturbation Theory, the critical
exponents are given by mean-field theory even at next-to-leading
order. From the quality of the fits of the lattice results with Chiral
Perturbation Theory that are presented above, we conclude that the
critical exponents measured on the lattice are consistent with mean field.

We notice that Chiral Perturbation Theory at leading order does not describe
the data very well at large chemical potential and diquark source. This is
presumably a sign that higher order corrections have to be taken into
account. The introduction of $\mu$ breaks the original symmetry and allows the
condensates to vary independently. This can, in part, be implemented by
replacing the effective potential based on non-linear sigma models, by a
phenomenological effective potential suggested by linear sigma models, where
the condensates are less tightly constrained \cite{KS_iso}.

\section{Scaling function}

The static scaling hypothesis states that the data for $\langle \chi^T \tau_2
\chi \rangle/(\mu-\mu_c)^{\beta_{mag}}$ for different diquark sources and
chemical potentials should collapse onto a single curve for $\mu>\mu_c$ when
plotted against $\lambda/(\mu-\mu_c)^\Delta$ \cite{Goldenfeld}. The gap
exponent $\Delta=\beta_{mag} \delta$; $\Delta=3/2$ in mean-field theory. We
can fit the lattice results to the scaling function if we assume some form for
the scaling function. We will try several possibilities for the form of the
scaling function: linear, of the form $f(x)=a +b x^s$, of the form
$f(x)=a+b x+cx^{2/3}$ and the simplest mean-field
form. The number of parameters changes according to the assumption for the form
of the scaling function. We present our results in Table~4 using part of the
data ($\mu=0.297$, $0.30$, $0.31$, $0.32$, $0.33$, and $0.34$).

\begin{center}
\begin{figure}[htb!]
\vspace{.5cm}
\hspace*{-.1cm}
\epsfig{file=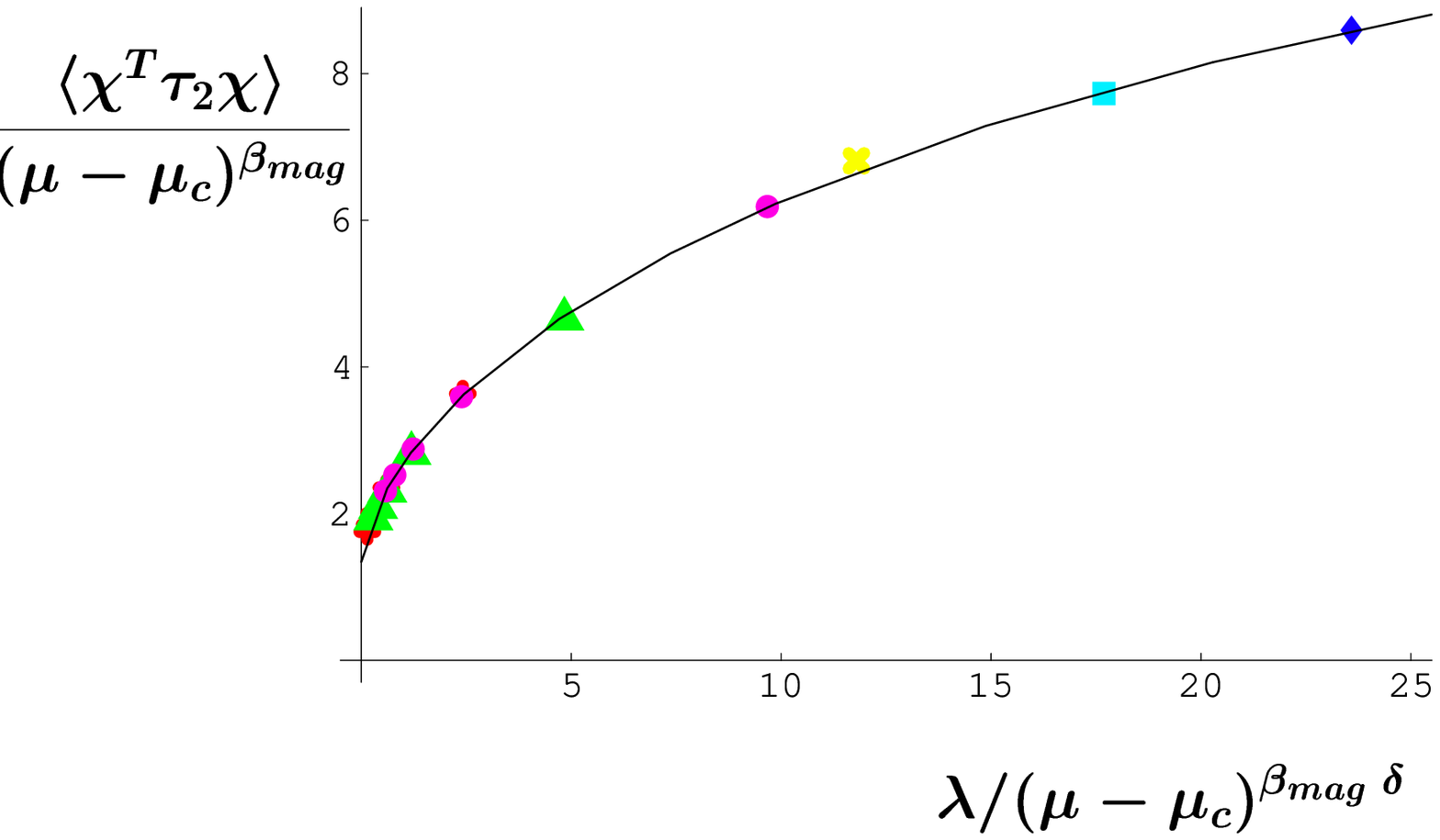, width=13.3cm, height=8cm}\vspace{2cm}
\epsfig{file=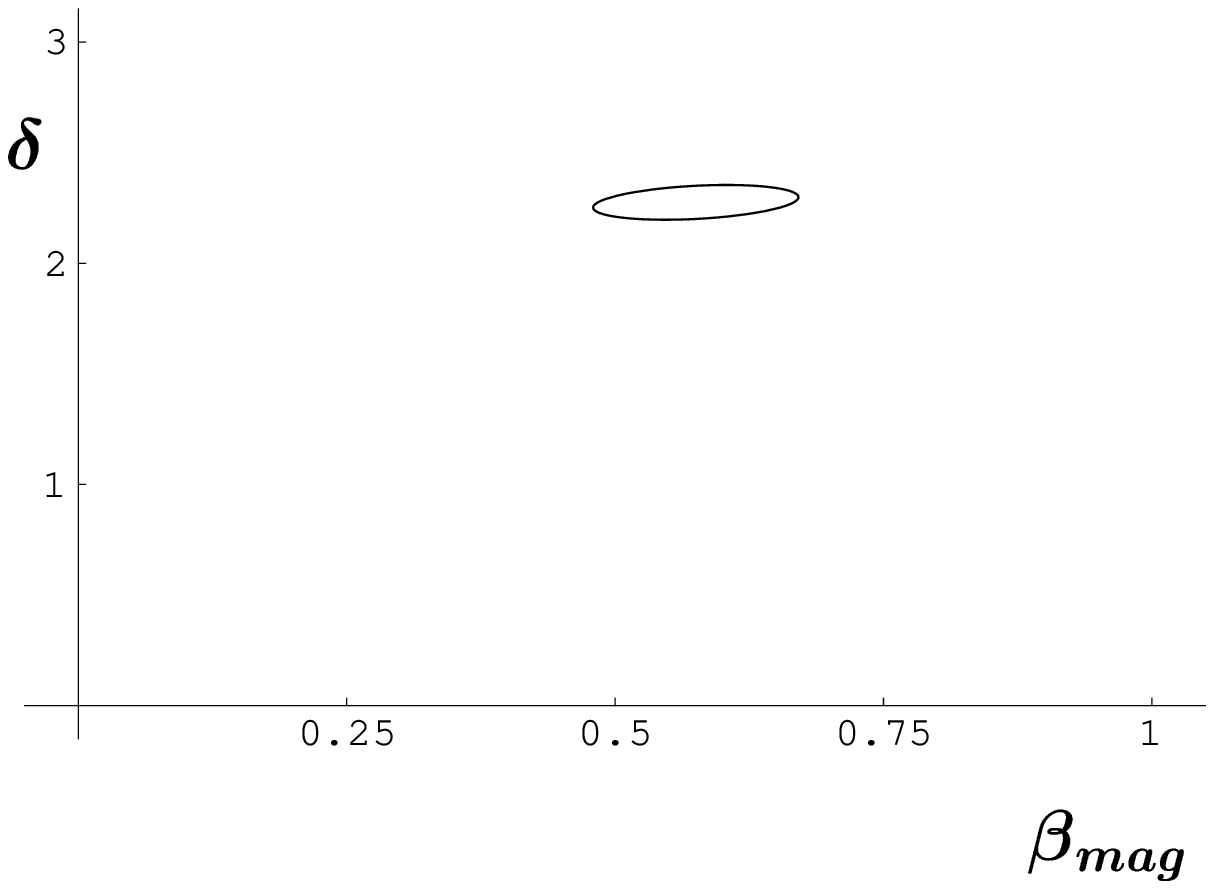, width=13.3cm, height=8cm}\vspace{.5cm}
\caption[]{\small Scaling function and $95\%$-confidence ellipse for
the critical exponents $\beta_{mag}$ and 
$\delta$ from our best fit, which corresponds to $\mu_c=0.2948$,
$\beta_{mag}=0.58$, and $\delta=2.28$. The pluses are the 
lattice data for $\lambda=0.025$, the xs for $\lambda=0.004$,
the triangles for $\lambda=0.005$, the squares for
$\lambda=0.006$,  the diamonds for $\lambda=0.008$, and the
dots for 
$\lambda=0.010$.}
\label{fig:scale}
\end{figure}
\end{center}

The different forms assumed for the scaling function give similar results for
the critical chemical potential and for the two critical exponents that we
studied. The best fit corresponds to a critical chemical potential that is
consistent with the result we found in the previous section,
$\mu_c=0.2948(7)$, a value of $\beta_{mag}=0.58(4)$ which is consistent with
mean field, and a value of $\delta=2.28(3)$ which is about three quarters of
the mean-field result. None of these fits are very good: the best fit has a
$\chi^2/dof=17$. It corresponds to the scaling function of the form
$f(x)=a+b x+cx^{2/3}$, as presented in Table~VI. In
Figure~\ref{fig:scale}, we show the scaling function 
together with the $95\%$-confidence ellipse for the parameters $\beta_{mag}$
and $\delta$ using the best fit.

The quality of these fits is rather poor. Furthermore, we find that there is a
sizeable correlation between the different fit parameters: $\mu_c$ is strongly
anti-correlated with $\beta_{mag}$ and mildly with $\delta$,  whereas
$\beta_{mag}$ and $\delta$ are mildly correlated with each other. Therefore,
the results from the fits using the scaling function are somewhat suspect.
This can be readily seen by using the same data as above plus the four points
that correspond to $\mu=0.2920$. The best fit we get has a $\chi^2/dof=66$. It
gives $\mu_c=0.269(8)$, $\beta_{mag}=0.7(1)$, and $\delta=1.4(2)$. Notice that
$\mu_c$ is much lower than before, that the increase in $\beta_{mag}$ is
sizeable, and that $\delta$ has changed by almost $40\%$.  This result
illustrates the problems usually encountered in the use of the scaling
function. Since the scaling function is in general different above and below
the critical point, the choice of data is crucial, and it is difficult to get a
clear result from this type of analysis. The comparison with Chiral
Perturbation Theory presented in the previous section does not suffer from
these problems.

Finally if we impose mean-field exponents, using the same part of
the data as above ($\mu=0.297$, $0.30$, $0.31$, $0.32$, $0.33$, and
$0.34$), the best fit we get
has poor quality: $\chi^2/dof=243$. It gives $\mu_c=0.297(8)$, which
is compatible with the results obtained in the previous section where
we compared the data with Chiral Perturbation Theory. If we try to use
scaling functions derived from the mean-field equation of state as in
\cite{KS_iso}, the quality of the fit is marginally
better, with a 
$\chi^2/dof=205$, and it gives $\mu_c=0.308(2)$. 

%\subsection{Extrapolation to $\lambda=0$}
%
%For each chemical potential $\mu=0.30$, $0.31$, $0.32$, $0.33$, and
%$0.34$, we first extrapolate the 
%diquark condensate to zero diquark source from its value at $\lambda=0.0025$,
%$\lambda=0.050$, and $\lambda=0.010$. We
%use a power extrapolation. From the extrapolation we can determine
%the critical chemical potential 
%and $\beta$. We fit the extrapolated diquark condensate to $a
%(\mu-\mu_c)^\beta+b (\mu-\mu_c)^s$. 
%We get that $\mu_c=0.29\pm0.04$ and that
%$\beta_{mag}=0.51\pm0.06$. The fit is good: $R^2=0.99998$. The 
%results of this Section are consistent with mean field.
%
%\subsection{Moral}
%
%With these three methods, it seems clear that $\beta_{mag}$ should
%have its mean-field value, and that a 
%very safe range for the critical chemical potential is
%$\mu_c=0.30\pm0.02$. On the other hand, 
%$\delta$ is hard to determine. We don't have any clear-cut argument to 
%determine its value.
%We cannot, however, exclude the mean-field value $\delta=3$.

\section{Runs at finite temperature}

We now present our $12^3 \times 6$ simulations to investigate
the interior of the phase diagram of Fig.~\ref{fig:phase} . 
We are particularly interested in scanning the diagram starting at
$\mu=0$ and following phase transitions 
and lines of crossover phenomena at high $T$ out to large $\mu$. In
particular, the low mass four flavor
theory should have a first order transition
on the $\mu=0$ axis and this transition should 
penetrate into the phase diagram. In addition, the $T=0$ transition to
a diquark condensate at $\mu_c$ should 
penetrate into the phase diagram and separate the normal
phase from the phase with a diquark condensate
as shown in Fig.~1. It
would be interesting to understand if and how these two lines of transitions
merge inside the phase diagram.
Finally, we know from past simulations and Effective Lagrangian
analyses that there is a line of first order 
transitions at high $\mu$ and high $T$ which separates the diquark
condensate phase from the quark-gluon plasma 
phase \cite{PLB}. We want to confirm the first order character of the
high $\mu$ part of this line and to find the low $\mu$ end of this
line where it changes to second order at a tricritical point, labeled
2 in Fig.~\ref{fig:phase}. This point has also been found within
Chiral Perturbation at nonzero temperature and chemical potential
\cite{STV2}. 

We are hopeful that some of these features of the phase diagram will
also occur in other models, so it would be 
useful to confirm them in this relatively simple setting. Perhaps, QCD
at nonzero Baryon Number chemical potential 
has a $T$-$\mu$ phase diagram with some of these features.

\begin{figure}[htb!]
\centerline{
\epsfxsize 5 in
\epsfysize 3 in
\epsfbox{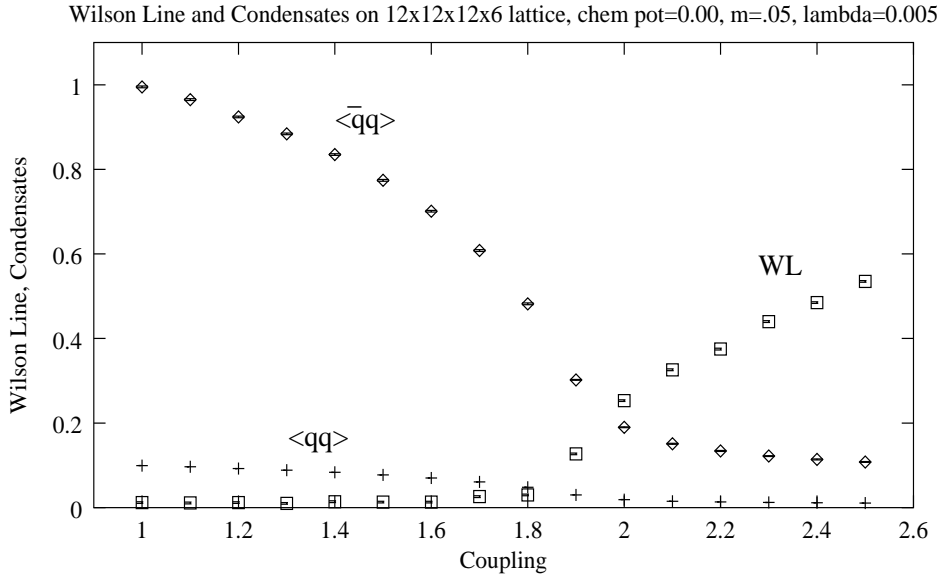}
}
\caption[]{Wilson Line and Condensates vs. $\beta$ for $\mu=0.00$.}
\label{fig:mu=0}
\end{figure}

We begin by considering the $\mu=0$ axis. Our simulations are at nonzero
fermion mass $m=0.05$ and non-zero diquark source $\lambda=0.005$
(equivalent to running at $m=\sqrt{0.05^2 + 0.005^2}$ ). $m=0.05$ is large 
enough that we expect the transition between hadronic matter and the
quark-gluon plasma to be 
weakened and perhaps softened into a 
crossover. This is, in fact, what we find. (At lower quark mass we would expect
to find a first order transition.)
In Fig.~\ref{fig:mu=0} we show data
for the chiral condensate and the Wilson Line 
and confirm a smooth but rapid crossover in the vicinity of $\beta=1.9$.

\begin{figure}[htb!]
\centerline{
\epsfxsize 5 in
\epsfysize 3 in
\epsfbox{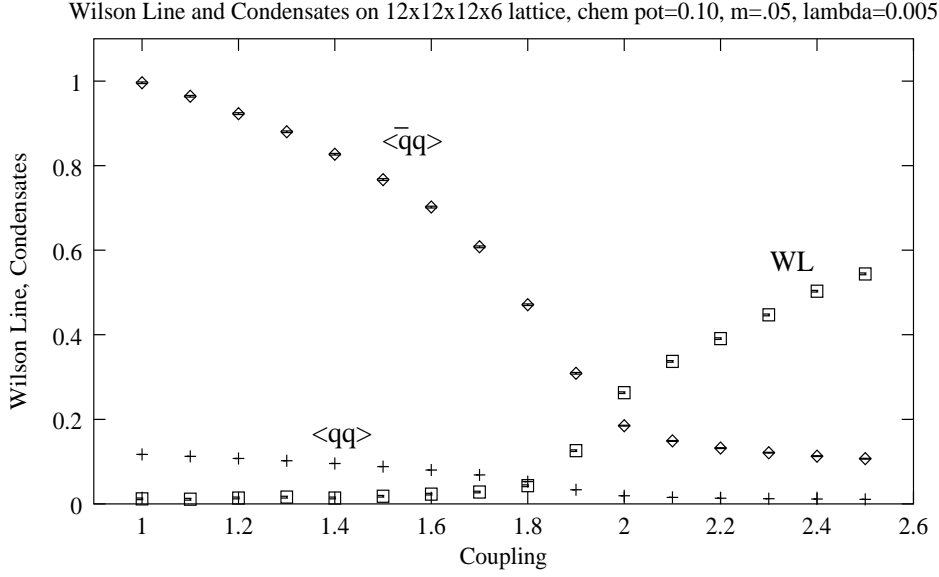}
}
\caption[]{Wilson Line and Condensates vs. $\beta$ for $\mu=0.10$.}
\label{fig:mu=0.1}
\end{figure}

Measurements at $\mu=0.10$ gave similar conclusions: there is a clear
but smooth crossover in the vicinity of $\beta=1.9$, as shown in 
figure~\ref{fig:mu=0.1}. 

\begin{figure}[htb!]
\centerline{
\epsfxsize 5 in
\epsfysize 3 in
\epsfbox{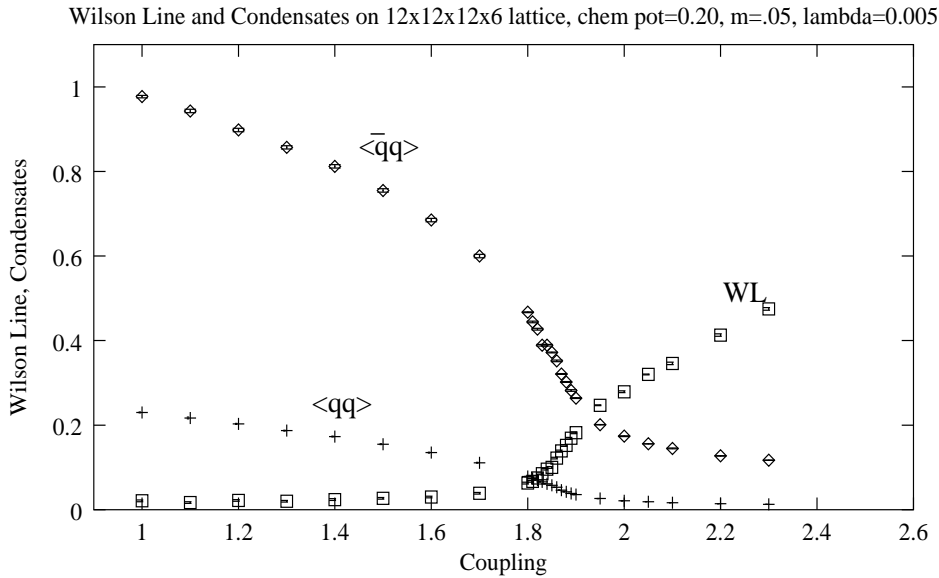}
}
\caption[]{Wilson Line and Condensates vs. $\beta$ for $\mu=0.20$.}
\label{fig:mu=0.2}
\end{figure}

It is interesting that a plot of the same quantities at $\mu=0.20$ displays a
noticeably sharper crossover.
Figure~\ref{fig:mu=0.2} suggests that $\mu=0.20$
is near a critical point, such as the point 1
in our generic phase diagram, Fig.~\ref{fig:phase}. At this rather large quark
mass $m=0.05$, this appears merely to be a more pronounced crossover
but, perhaps, if $m$
were chosen smaller than $0.05$, 
an actual line of first order transitions would be found for the region 
corresponding to $\mu>0.20$ in this phase diagram.

\begin{figure}[htb!]
\centerline{
\epsfxsize 5 in
\epsfysize 3 in
\epsfbox{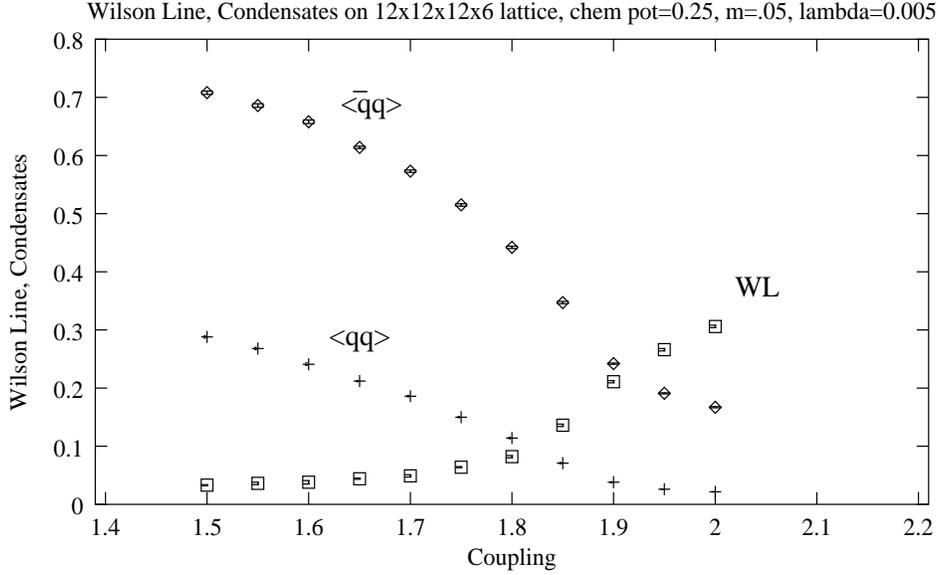}
}
\caption[]{Wilson Line, and Condensates vs. $\beta$.}
\label{fig:mu=0.25}
\end{figure}

Once we increase $\mu$ to $0.25$, the induced diquark condensate
becomes more significant at low $T$, as shown in
figure~\ref{fig:mu=0.25}.

\begin{figure}[htb!]
\centerline{
\epsfxsize 5 in
\epsfysize 3 in
\epsfbox{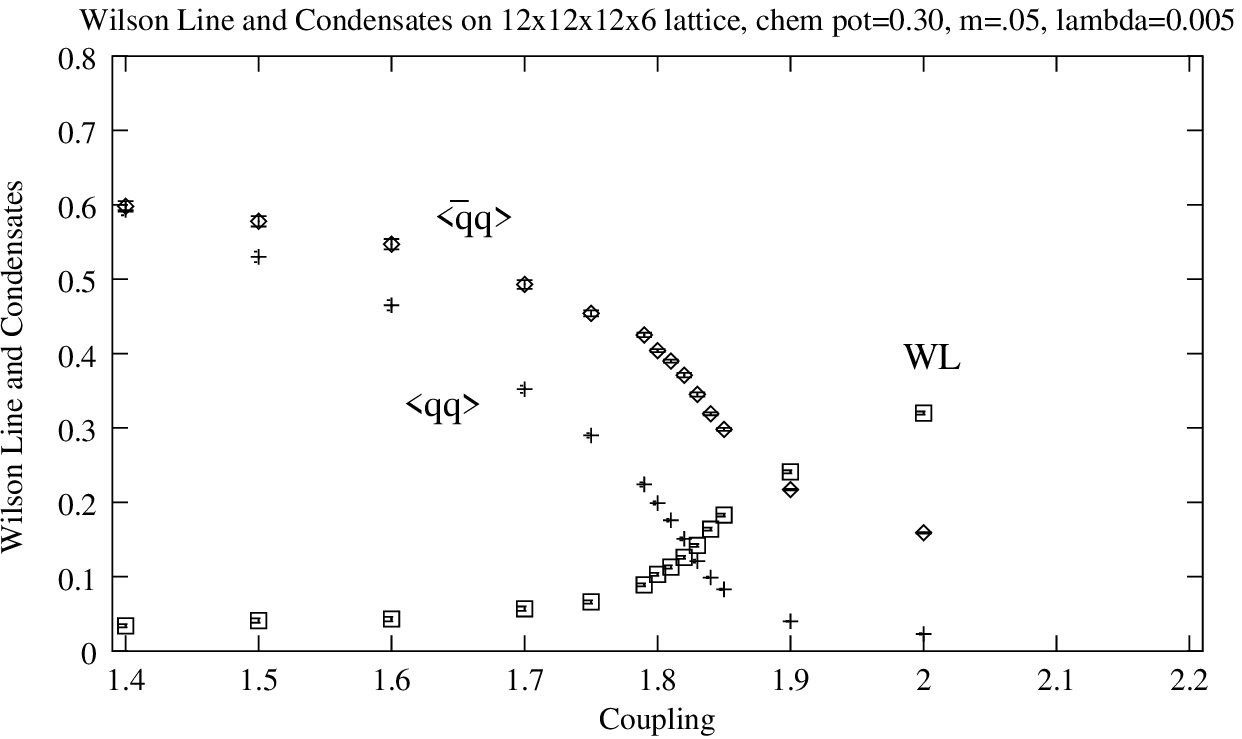}
}
\caption[]{Wilson Line and Condensates vs. $\beta$.}
\label{fig:mu=0.3}
\end{figure}

At $\mu=0.30$ the diquark condensate becomes the most interesting and
rapidly varying quantity. The chiral 
condensate is relatively smooth and changes from $0.493(6)$ at
$\beta=1.7$ to $0.217(1)$ at $\beta=1.9$ 
while the diquark condensate has varied from $0.352(5)$ to $0.0400(3)$
over the same range of coupling. The 
Wilson Line varies from $0.057(3)$ to $0.241(2)$ over the same
range. As shown in the figure~\ref{fig:mu=0.3}, the 
diquark condensate experiences a sharp transition between $\beta=1.80$
and $1.90$. 

\begin{figure}[htb!]
\centerline{
\epsfxsize 5 in
\epsfysize 3 in
\epsfbox{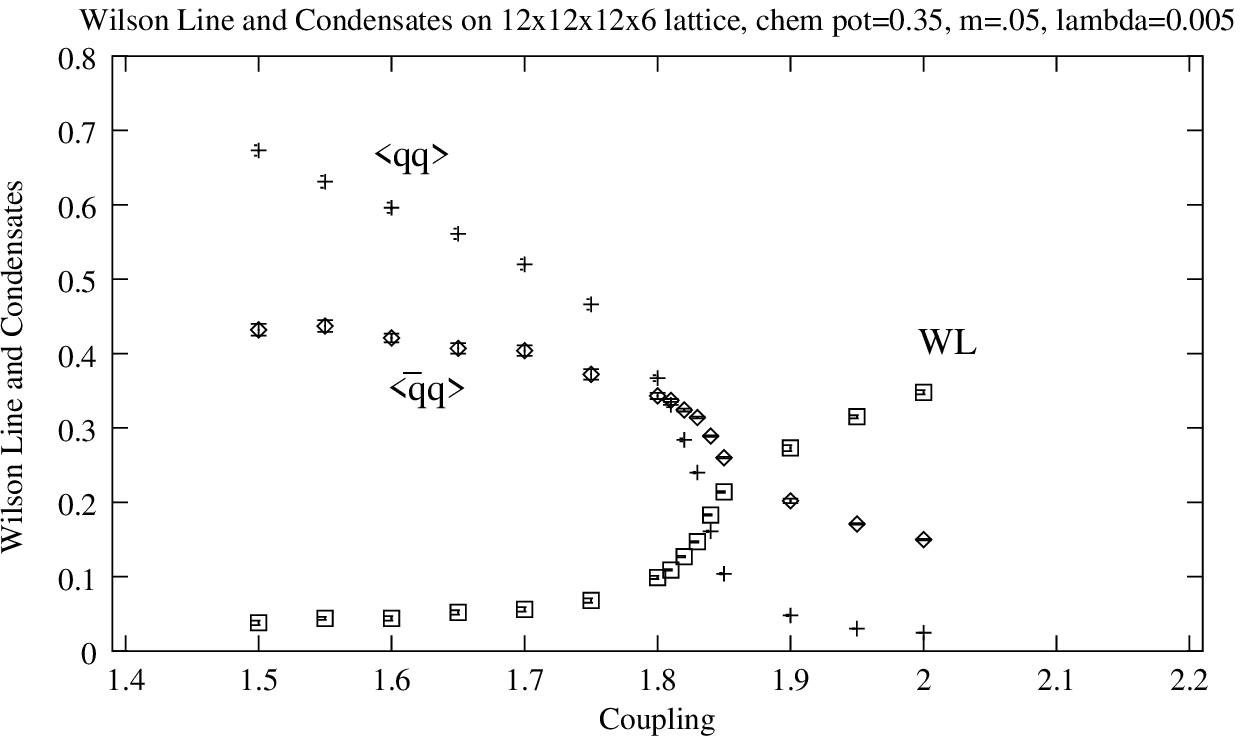}
}
\caption[]{Wilson Line and Condensates vs. $\beta$.}
\label{fig:mu=0.35}
\end{figure}

That transition becomes even sharper as we increase $\mu$ to $0.35$ as
shown in figure~\ref{fig:mu=0.35}. 

\begin{figure}[htb!]
\centerline{
\epsfxsize 5 in
\epsfysize 3 in
\epsfbox{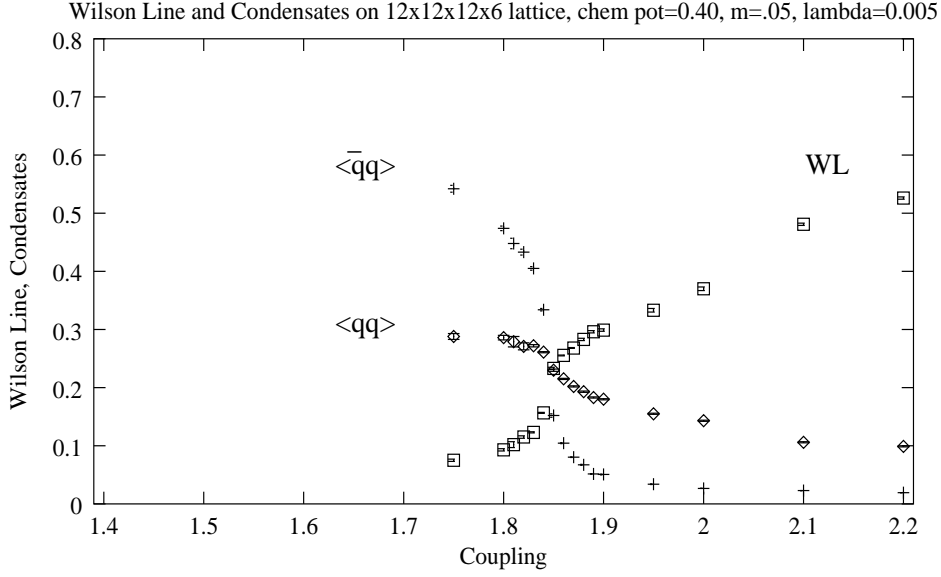}
}
\caption[]{Wilson Line and Condensates vs. $\beta$.}
\label{fig:mu=0.4}
\end{figure}

At $\mu=0.40$, the simulations indicate a first order transition near
$\beta=1.85(3)$. In figure~\ref{fig:mu=0.4} 
we show both the diquark condensates and the Wilson Line and see
strong suggestions of discontinuities. 

\begin{figure}[htb!]
\centerline{
\epsfxsize 5 in
\epsfysize 3 in
\epsfbox{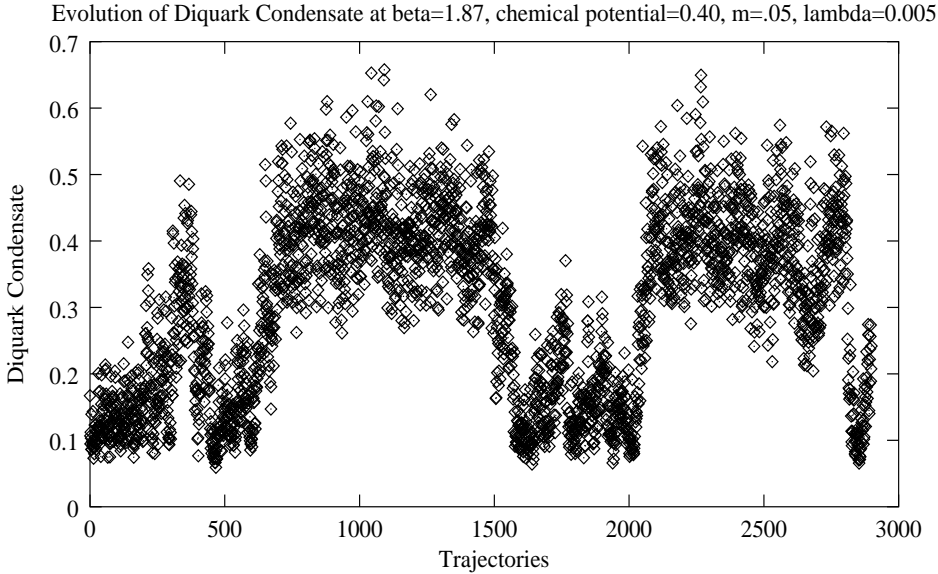}
}
\caption[]{Diquark Condensate vs. Computer Time.}
\label{fig:meta}
\end{figure}

The really solid evidence for a first order transition comes from the
time evolution of the observables 
at $\beta=1.87$ which show signs of metastability. For example, in the
figure~\ref{fig:meta} we show the time evolution 
of the diquark condensate and display several tunnelings between a
state having a condensate 
near $0.15$ and another with a condensate near $0.40$.

\begin{figure}[htb!]
\centerline{
\epsfxsize 5 in
\epsfysize 3 in
\epsfbox{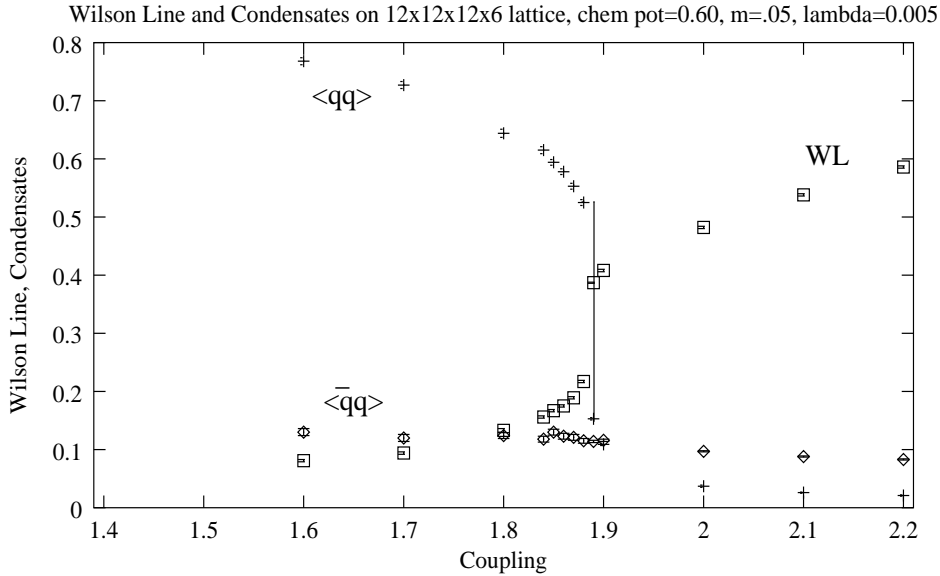}
}
\caption[]{Wilson Line and Condensates vs. $\beta$.}
\label{fig:mu=0.6}
\end{figure}

Runs at $\mu=0.50$ and $0.60$ show clear discontinuities in both the
diquark condensate and the Wilson 
Line near $\beta=1.87$. In figure~\ref{fig:mu=0.6} we show the results for
$\mu=0.60$.

\begin{figure}[htb!]
\centerline{
\epsfxsize 5 in
\epsfysize 3 in
\epsfbox{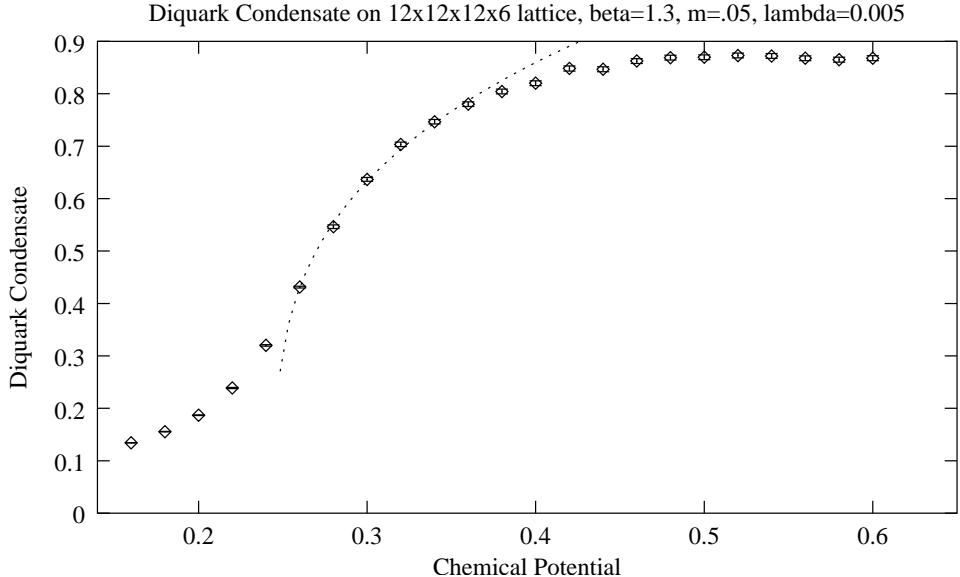}
}
\caption[]{Diquark Condensate  vs. $\mu$.}
\label{fig:beta=1.3}
\end{figure}

Finally, we scanned the phase diagram at low $T$ and variable
$\mu$. We accumulated data at 
$\beta=1.30$, relatively strong coupling and near vanishing $T$, on
the $12^3 \times 6$ lattice 
and took measurements over a wide range of $\mu$, from $0.15$ to
$0.60$. The results for the diquark 
condensate are shown in the figure~\ref{fig:beta=1.3} which shows a transition
near $\mu=0.25$. 
The power law fit shown in the figure produces a critical index
$\beta_{mag}=0.29(4)$ at a critical chemical potential 
$\mu_c=0.2453(2)$. The quality of the fit, which extends from
$\mu=0.26$ to $0.38$, is rather poor, having 
a confidence level of only a fraction of one percent. As we have seen in 
quenched simulations and in simulations of QCD at finite chemical
potential for isospin such small estimates of the critical index are often an
indication that the scaling is best described in terms of the scaling form
for the effective Lagrangian in the non-linear sigma-model class.
\cite{KS_iso,KS_iso2,KLS}.

The estimate for the critical chemical potential, $\mu_c=0.2453(2)$, meshes
well with our measurements at high $T$ above, which indicated that the
diquark condensate only started to be numerically significant for $\mu \gtrsim
0.25$.

\begin{figure}[htb!]
\centerline{
\epsfxsize 5 in
\epsfysize 3 in
\epsfbox{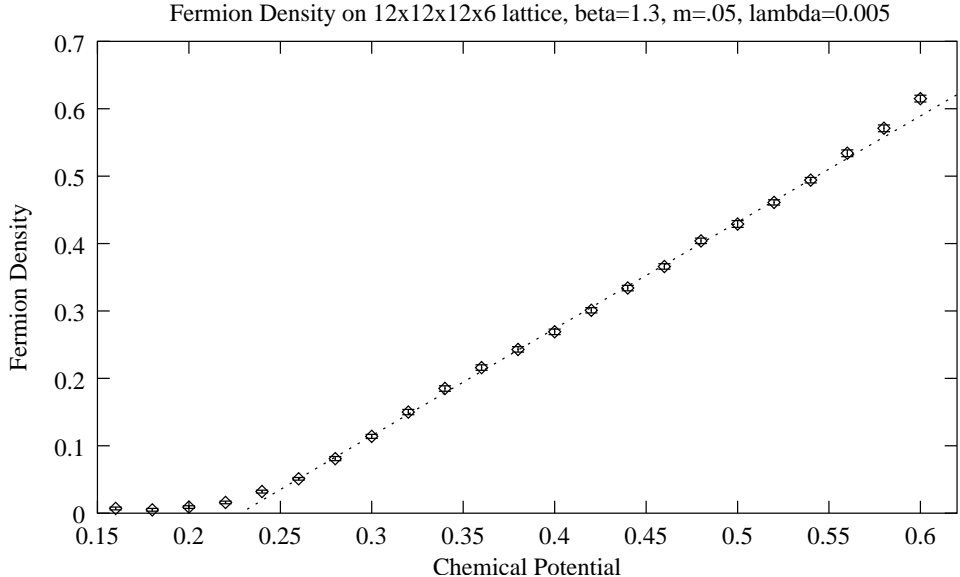}
}
\caption[]{Fermion-Number Density  vs. $\mu$.}
\label{fig:j0b13}
\end{figure}

The continuous phase transition is also seen in the induced fermion-number
density shown in the figure~\ref{fig:j0b13}. The dotted 
line fit to that data, which extends all the way from $\mu=0.26$ to $0.56$,
has a confidence level of $42$ percent, a critical index $0.99(2)$ and an
estimate for the critical chemical potential, $\mu_c=.2282(2)$. The scaling
law obtained here is in good agreement with field theory expectations of
unity. Note that such linearity for these relatively strong couplings has
a wider range of validity than would be indicated by the scaling window, as
is seen in the explicit scaling forms from effective Lagrangians~\cite{Toublan,%
SUNY}. Since we only have a single $\lambda$ value, we have not attempted a
fit to such a form.
%The agreement is good, presumably because it is controlled by current
%conservation, which applies at strong coupling as well as weak, in the scaling
%window.

%BREAK

\section{Conclusion and Outlook}

In this article we have presented a study of the phase diagram of
2-color QCD at nonzero baryon/quark-number chemical potential and nonzero
temperature. We have found two phases: a ``normal'' phase where quark
number is conserved and a phase with a diquark condensate which spontaneously
breaks quark number. At $\mu=0$, for smaller quark masses, there should be a first
order finite temperature ``deconfinement'' transition which divides the
hadronic from the quark-gluon plasma phase. Since, even in this case we
expect the line of first order transitions emerging from this point to
terminate at small $\mu$, beyond which there should be a rapid crossover but
no transition, the hadronic and quark-gluon plasma phases are not distinct.
This is the phase we call the ``normal'' phase. The chiral condensate is
non-zero everywhere. However, where it is large the system is more hadronic
in nature; where it is small, the system is more plasma-like. In the ``diquark''
phase, the chiral condensate decreases with increasing diquark condensate,
approaching zero for large $\mu$ (even before saturation sets in). 
We have found that the phase transition
between the ``normal'' phase and the diquark phase for small $T$ is second
order and compatible with mean field theory. 
%At zero temperature, the critical
%chemical potential is given by half the zero-$T$ pion mass. HOW DO YOU KNOW --
%WE DIDN'T MEASURE THE PION MASS, DID WE? 
These simulations therefore agree with the predictions of Chiral
Perturbation Theory through next-to-leading order \cite{STV1}.

We have also found that the second order phase transition between the
``normal'' phase and the diquark phase becomes first order when $\mu$
and $T$ are increased. Therefore the $\mu$-$T$ phase diagram contains
a tricritical point. It corresponds to a chemical potential near two-thirds of
the pion mass, and a temperature only slightly lower than the transition
temperature between the hadronic phase and quark-gluon plasma phase at
$\mu=0$. The tricritical point is also found in Chiral Perturbation Theory at
nonzero temperature and chemical potential \cite{STV2}. Therefore,
this phase diagram is consistent with the one obtained within Chiral
Perturbation Theory \cite{STV2}.

Now that the phase diagram of this theory has been mapped out, we can turn to
more quantitative properties. In particular, we plan on analyzing the gauge
configurations for their topological content. There are interesting
predictions \cite{SSZ} based on Effective Lagrangians for the size, density
and interactions among the instantons of the model at large $\mu$ which can be
investigated by cooling methods which have been very successful at vanishing
$\mu$. In addition, we can simulate the phase diagram at low $\mu$ and high T
with lighter quarks and search for the critical point 1. The SU(3) analog
of this point is thought to be accessible to heavy ion collisions planned for
BNL's RHIC.

Furthermore, the spectroscopy of the model, with an emphasis on Goldstone and
pseudo-Goldstone bosons, is also under consideration. Since the theory's light
modes control its critical behavior and thermodynamics, quantitative results
on spectroscopy are quite important.

There are also interesting effective Lagrangian predictions for QCD with
chemical potentials associated with the light quarks \cite{SS,KT,SSSTV}. We
are studying QCD with a chemical potential associated with isospin
\cite{KS_iso2,STV2}. This model's phases at nonzero $\mu$ and temperature $T$ are
expected to be very similar to those discussed here \cite{SS}. Although we
cannot attack the $SU(3)$ theory with a large Baryon number chemical
potential, these other situations can be studied both analytically and
numerically, and interesting new phases of matter have been found there which
should be investigated further.

\section*{Acknowledgment}
This work was partially supported by NSF under grant NSF-PHY-0102409
and by the U.S. Department of Energy under contract W-31-109-ENG-38.
D.T. is supported in part by ``Holderbank''-Stiftung.
The simulations were done at NPACI and NERSC. B. Klein,
K. Splittorff, M.A. Stephanov and J. Verbaarschot are acknowledged for useful
discussions.

\begin{table}
\begin{tabular} {cddd}
$\mu$ & $<\chi^T\tau_2\chi>$ & $<\chi^T\tau_2\chi>$  & $<\chi^T\tau_2\chi>$  \\
\hline
$.25$ &   0.0617(2)  &  0.1134(2)       & 0.1962(4)  \\
$.28$ &   0.1120(5)  &  0.1732(4)       & 0.2546(5)  \\
$.29$ &   0.1407(5)  &  0.1993(5)       & 0.2782(4)  \\
$.30$ &   0.1772(9)  &  0.2269(6)       & 0.3016(5)  \\
$.31$ &   0.2120(6)  &  0.2544(6)       & 0.3240(3)  \\
$.32$ &   0.2415(11) &  0.2783(7)       & 0.3470(6)  \\
$.33$ &   0.2711(7)  &  0.3045(7)       & 0.3687(5)  \\
$.34$ &   0.2951(11) &  0.3253(9)       & 0.3893(6)  \\
$.35$ &   0.3200(9)  &  0.3499(7)       & 0.4082(7)  \\
$.36$ &   0.3437(12) &  0.3651(9)       & 0.4287(6)  \\
$.38$ &   0.3815(13) &  0.4044(9)       & 0.4627(7)  \\
\end{tabular}
\caption{ Diquark Condensates on a $16^4$ lattice at $\lambda=0.0025$,
$0.005$, and $0.010$ in the second, third, and fourth column,
respectively. The first column 
lists the $\mu$ values.}
\end{table}

\begin{table}
\begin{tabular} {cddd}
$\mu$    & $<\chi^T\tau_2\chi>$ & $<\chi^T\tau_2\chi>$    \\ \hline
$.25$ &   0.0100(2)  &  0.0031(2)      \\
$.28$ &   0.0508(5)  &  0.0371(5)      \\
$.29$ &   0.0821(5)  &  0.0693(5)      \\
$.30$ &   0.1275(9)  &  0.1193(9)      \\
$.31$ &   0.1710(6)  &  0.1664(6)      \\
$.32$ &   0.2050(11) &  0.2031(11)      \\
$.33$ &   0.2380(7)  &  0.2376(7)      \\
$.34$ &   0.2650(11) &  0.2661(11)      \\
$.35$ &   0.2900(9)  &  0.2896(9)      \\
$.36$ &   0.3220(12) &  0.3292(12)      \\
$.38$ &   0.3590(13) &  0.3628(12)      \\
\end{tabular}
\caption{Diquark Condensates on a $16^4$ lattice at $\lambda=0.00$. The first
column lists the $\mu$ values, the second records the linear extrapolations of
the diquark condensates and the third records the quadratic extrapolations.}
\end{table}

\begin{table}
\begin{tabular} {cddd}
$\mu$ & $<\bar{\chi}\chi>$ & $<\bar{\chi}\chi>$  & $<\bar{\chi}\chi>$   \\ \hline
$.25$ & 0.4123(5)    & 0.4012(5)       & 0.3887(5)   \\
$.28$ & 0.4025(8)    & 0.3850(6)       & 0.3697(6)  \\
$.29$ & 0.3934(7)    & 0.3758(5)       & 0.3616(4)  \\
$.30$ & 0.3816(9)    & 0.3664(7)       & 0.3529(5)  \\
$.31$ & 0.3686(6)    & 0.3541(6)       & 0.3444(5)  \\
$.32$ & 0.3523(11)  & 0.3409(8)       & 0.3335(6)  \\
$.33$ & 0.3381(6)    & 0.3298(8)       & 0.3234(5)  \\
$.34$ & 0.3242(11)  & 0.3157(8)       & 0.3125(6)  \\
$.35$ & 0.3091(10)  & 0.3057(7)       & 0.3026(7)  \\
$.36$ & 0.2987(12)  & 0.2908(9)       & 0.2920(7)  \\
$.38$ & 0.2733(12)  & 0.2722(9)       & 0.2711(7)  \\
\end{tabular}
\caption{ Quark-antiquark Condensates on a $16^4$ lattice at $\lambda=0.0025$,
$0.005$, and $0.010$ in the second, third, and fourth column,
respectively. The first column records the $\mu$ values.}
\end{table}

\begin{table}
\begin{tabular} {cdddd}
$\lambda$    & $<\chi^T\tau_2\chi>$ & $<\chi^T\tau_2\chi>$  &
$<\chi^T\tau_2\chi>$ & $<\chi^T\tau_2\chi>$  \\ \hline 
$.002$ &   0.1239(6)  &  0.1315(4)     & 0.1678(8)  &  0.2194(14)    \\
$.004$ &   0.1678(4)  &  0.1732(3)     & 0.1890(5)  &  0.2034(7)     \\
$.006$ &   0.2031(4)  &  0.2073(2)     & 0.2190(4)  &  0.2308(5)     \\
$.008$ &   0.2303(4)  &  0.2338(2)     & 0.2440(4)  &  0.2566(4)     \\
$.010$ &   0.2526(3)  &  0.2579(2)     & 0.2662(3)  &  0.2753(4)     \\
\end{tabular}
\caption{ Diquark Condensates on a $16^4$ lattice at four estimates of
the critical $\mu$, $0.2855$, $0.2870$, 
$0.2920$, and $0.2970$ in the second, third, fourth, and fifth column,
respectively. The first column records the $\lambda$ values.}
\end{table}

\begin{table}
\begin{tabular} {cdddd}
$\lambda$    & $<\bar{\chi}\chi>$ & $<\bar{\chi}\chi>$  &
$<\bar{\chi}\chi>$ & $<\bar{\chi}\chi>$  \\ \hline 
$.002$ &  0.4244(9) &  0.4261(6)   &  0.4362(9)   &  0.452(2)  \\
$.004$ & 0.3882(6)  & 0.3875(4)  & 0.3836(5) & 0.3780(9)  \\
$.006$ & 0.3730(4)  & 0.3726(3)  & 0.3668(4) & 0.3619(5)  \\
$.008$ & 0.3622(4)  & 0.3611(3)  & 0.3566(4) & 0.3511(4)  \\
$.010$ & 0.3525(4)  & 0.3528(2)  & 0.3464(3) & 0.3415(4)  \\
\end{tabular}
\caption{ Quark-antiquark Condensates on a $16^4$ lattice at four estimates of
the critical $\mu$, $0.2855$, $0.2870$, 
$0.2920$, and $0.2970$ in the second, third, fourth, and fifth column,
respectively. The first column records the $\lambda$ values.}
\end{table}

%\begin{table}[h]
%\centering
%\caption{\footnotesize{Comparison between ChPT and the lattice results using 
%the whole
%data set ($\mu=0.25$, $0.28$, $0.30$, $0.31$, $0.32$, $0.33$, $0.34$,
%$0.35$, $0.36$, and $0.38$). 
%The critical chemical potential, two critical exponents, and a measure of the 
%quality of the
%one-parameter fits are given.}}
%\begin{tabular}{|c|c|c|c|c|}
%\hline
% & $\lambda=0.0025$ & $\lambda=0.005$ & $\lambda=0.010$ & all diquark sources
% together
%\hline
%$\mu_c$ & $0.300\pm0.002$ & $0.298\pm0.002$ & $0.294\pm0.003$ & 
%$0.298\pm0.002$\\
%$\beta$ & $1/2$ & $1/2$ & $1/2$ & $1/2$ \\
%$\delta$ & $3$ &  $3$ &  $3$ &  $3$ \\
%$R^2$ & $0.99899$ & $0.99907$ & $0.99872$ & $0.99816$ \\
%\hline
%\end{tabular}
%\end{table}
%
%\begin{table}[h]
%\centering
%\caption{\footnotesize{Comparison between ChPT and the lattice results using 
%part of the
%data ($\mu=0.30$, $0.31$, $0.32$, $0.33$, and $0.34$). The critical chemical 
%potential,
%two critical exponents, and a measure of the quality of the one-parameter fits
% are given.}}
%\begin{tabular}{|c|c|c|c|c|}
%\hline
% & $\lambda=0.0025$ & $\lambda=0.005$ & $\lambda=0.010$ & all diquark sources 
%together \\
%\hline
%$\mu_c$ & $0.302\pm0.001$ & $0.300\pm0.002$ & $0.296\pm0.003$ & 
%$0.300\pm0.001$\\
%$\beta$ & $1/2$ & $1/2$ & $1/2$ & $1/2$ \\
%$\delta$ & $3$ &  $3$ &  $3$ &  $3$ \\
%$R^2$ & $0.99991$ & $0.99979$ & $0.99971$ & $0.99910$ \\

%\end{tabular}
%\end{table}

\begin{table}
\centering
\begin{tabular}{|c|c|c|c|}
\hline
Form of scaling function & $a+b x$ & $a+bx^s$ & $a+b x+c x^{2/3}$\\
\hline \hline
Number of parameters & 5 & 6 & 6 \\
\hline
$\mu_c$ & $0.278(7)$ & $0.285(7)$ & $0.2948(7)$ \\
$\beta_{mag}$ & $0.5(1)$ & $0.51(8)$ & $0.58(4)$ \\
$\delta$ & $1.1(2)$ &  $1.6(4)$ &  $2.28(3)$ \\
$\chi^2/dof$ & $86$ & $74$ & $17$ \\
\hline
\end{tabular}
\caption{\footnotesize{Determination of the critical chemical potential and
two critical exponents using the static scaling hypothesis assuming different
forms for the scaling function $f(x)$ and using the data for diquark sources
$0.0025\leq\lambda\leq0.010$, and $0.297\leq\mu\leq0.340$, i.e. 19 data
points. The numbers of parameters and a measure of the fit quality are given
for each fit.}}
\end{table}

\end{document}